# Probabilistic load flow calculation of AC/DC hybrid system based on cumulant method

Yinfeng Sun [a,*], Dapeng Xia [a,*], Zichun Gao [a], Zhenhao Wang [a], Guoqing Li [a], Weihua Lu [a], Xueguang Wu [b,c], Yang Li [a,*]

[a] *Key Laboratory of Modern Power System Simulation and Control & Renewable Energy Technology, Ministry of Education (Northeast Electric Power University), Jilin, 132012, China*
[b] *Global Energy Interconnection Research Institute, Beijing, 102211, China*
[c] *Beijing Key Laboratory of DC Power Grid Technologies and Simulation, Beijing, 102211, China*

**ABSTRACT:** The operating conditions of the power system have become more complex and changeable. This paper proposes a probabilistic load flow based on the cumulant method (PLF-CM) for the voltage sourced converter high voltage direct current (VSC-HVDC) hybrid system containing photovoltaic grid-connected systems. Firstly, the corresponding control mode is set for the converter, including droop control and master–slave control. The unified iterative method is used to calculate the conventional AC/DC flow. Secondly, on the basis of the probability model of load and photovoltaic output, based on the aforementioned flow results, use correlation coefficient matrix of this paper will change the relevant sample into independent sample, the cumulants of the load and photovoltaic output are obtained; Then, the probability density function (PDF) and cumulative distribution function (CDF) of state variables are obtained by using Gram-Charlie series expansion method. Finally, the mean value and standard deviation of node voltage and line power are calculated on the modified IEEE 34-bus and IEEE 57-bus transmission systems. The algorithm can reflect the inherent uncertainty of new energy sources, and replace the complex convolution operation, greatly improving the calculation speed and the convergence.

## 1. Introduction

In recent years, due to the advantages of low pollution and high reliability, the proportion of photovoltaic power generation in the power system has been increasing, which has become an increasingly important means to solve the current energy crisis. Because of its flexibility and reliability, voltage sourced converter high voltage direct current (VSC-HVDC) is well suited for renewable power integration and transmission. However, the photovoltaic output power has great fluctuation and uncertainty, and the output power of multiple power stations located close to each other has a certain correlation. Therefore, when calculating the random power flow of power systems, the influence of the correlation of the output of multiple photovoltaic power stations needs to be taken into account [1]. In addition, with the continuous growth of DC power supply and load in the power grid and the development of AC/DC hybrid power grid, the operation control of traditional power grids is facing severe challenges [2]. The traditional power flow calculation method is difficult to measure the influence of uncertain factors.

In 1974, Borkowaka proposed a probabilistic load flow (PLF) calculation method to solve many uncertain factors in a power system [3]. Since then, researchers all over the world have carried out a lot of research on this basis and achieved fruitful results. PLF solution methods can be divided into three categories: simulation method [4-6], approximation method [7-9] and analytical method [10,11]. The PLF calculation methods considering the correlation of input variables mainly include Monte Carlo simulation (MCS) method, point estimation method and convolution method. In [10,12], the author mentioned that MCS uses sampling technology to generate relevant samples, and then performs multiple deterministic power flow calculations, which can more accurately obtain the statistical distribution characteristics of output variables. However, a large number of simulation calculations lead to a long calculation time. It is often used as a reference to evaluate the advantages and disadvantages of other methods. The point estimation method approximates the statistical distribution characteristics of the output variables according to the digital characteristics of the input variables. The calculation speed is fast, but the high-order moment error of the output variables is large. Analytical methods include convolution method and cumulant method. In reference [13], the author mentioned that although the convolution method has a clear concept, it has a large amount of calculation and can only consider the linear correlation of input variables. The calculation involved in the stacking process using convolution is complex, difficult and time-consuming. By using the cumulant method, the convolution method can be transformed into a simple linear calculation. The fluctuation of new energy output and load can be used as disturbance variables. The probability distribution of node voltage and branch power flow can only be obtained by one conventional power flow calculation. The cumulant method has good prediction accuracy and high calculation efficiency in calculating PLF. However, the cumulant method requires each input variable to be independent of each other, so it can not be directly applied to the occasions where the input variables are related. However, due to the strong correlation between photovoltaic power stations close to each other, if the input variables are assumed to be independent of each other, there may be large errors in the probability distribution of state variables [14].

---

*Corresponding author.
   *E-mail addresses:* sunyinfeng@neepu.edu.cn (Y. Sun), 599905759@qq.com (D. Xia), liyang@neepu.edu.cn (Y. Li).

Although PLF calculation methods have been widely used in AC systems, grid operators also urgently need efficient and accurate PLF calculation methods for AC/DC hybrid systems to obtain new steady-state information of AC/DC hybrid distribution networks. However, there are few studies on probability methods in AC/DC hybrid systems at present, and MCS are mostly used. In [15], the author presents the probability flow model of AC/DC hybrid system under different control modes by using the node injection power operation curve to form the multiple operation mode of the system. The probability analysis of the DC hybrid system fails to consider the diversity of converter control methods. However, in actual AC/DC hybrid systems, the control mode of the inverter may change. At the same time, the current source converter used in this system is not suitable for forming a multi-node DC grid, so this method has certain limitations. In reference to the non-convergence of conventional power flow in PLF calculation, [16] solves these problems-accuracy, high-dimensionality, and computational time-with a coupled Karhunen-Loève (KL) expansion and Anisotropic Sparse Grid algorithm. This method has good robustness, but the calculation efficiency is low. In [17], a method of proportional scaling unscented transform for high-dimensional Gaussian distribution is proposed, which can accurately deal with the accuracy of high-dimensional PLF problem. Unscented transform belongs to approximation method. In order to improve the computational efficiency, approximate method and analytical method are proposed. Reference [18] proposed cluster the uncertain sources in the power system, and then calculate the AC/VSC-MTDC PLF according to the clustering center. Compared with the MCS calculation result, the calculation amount is reduced, and the obtained result has a certain degree of credibility, but the influence of the connection position of the renewable energy on the power distribution of the system is not considered. A PLF analysis is performed by modelling the variability of electric vehicle mobility, household load, photovoltaic system generation, and the adoption of photovoltaic system and electric vehicle in society in [19], while the research objective is to present electric vehicle smart charging schemes to increase the temporal matching between photovoltaic generation and the PLF algorithm only provides basic data for it, without detailed analysis.

In this paper, a probabilistic power flow based on cumulant method (PLF-CM) considering the correlation of input variables is proposed. In order to solve the problem that the cumulants of some input variables are difficult to be solved by traditional numerical methods, a method based on Monte Carlo sampling is proposed, which uses the samples of input variables to calculate their cumulants. The proposed algorithm solves the error problem of cumulant method in dealing with related random variables. By calculating the mean, variance and other statistical characteristics of state variables, PLF-CM can more accurately and quickly obtain the average value of node voltage and branch flow, so as to more comprehensively reflect the operation status of the AC/DC system.

Firstly, the AC/DC probabilistic power flow calculation model of the system is established in this paper and analyzes the influence of the different control methods of the converters at each end of the DC network on the model; then, the cumulant of photovoltaic power generation and load in the system is calculated, and combined with Gram-Charlier series expansion method to obtain the probability distribution curve of each node voltage and line power flow. Finally, the effectiveness of the proposed method is proved by simulation in the modified IEEE 34-bus and IEEE 57-bus systems.

## 2. Steady state mathematical model of AC/DC power grid

2.1. *DC power equation*

Multiple VSC converter stations are led out from the AC power grid, and the AC system is connected through multiple groups of point-to-point DC to form the AC and DC power grid. In the steady-state analysis, the DC network equation is expressed by the node voltage equation, i.e

$$I_{dk} = \sum_{j=1}^{n_{dc}} G_{dkj}(U_{dk} - U_{dj}) \tag{1}$$

Where: $I_{dk}$ is the injection current of the *k*-th dc node; $U_{dk}$ is the voltage of the *k*-th dc node; $G_{dkj}$ is $k$-$j$ conductance of DC line; $n_{dc}$ is the total number of DC grid nodes.

Consistent with the AC system, the active power $P_{dk}$ injected by each node of the DC network is usually known, i.e

$$P_{dk} = P_{dk,g} - P_{dj,d} \tag{2}$$

Where: $P_{dk,g}$ is the active power injection amount of DC power supply at the $k$-th dc node; $P_{dk,d}$ is the load active power at the $k$-th dc node.

Based on equations (1) and (2), the power equation of any node of DC power grid is shown in equation (3).

$$P_{dk} = U_{dk} \sum_{j=1}^{n_{dc}} G_{dkj}(U_{dk} - U_{dj}) \tag{3}$$

## 2.2. Converter station equivalent of AC involved dc node voltage

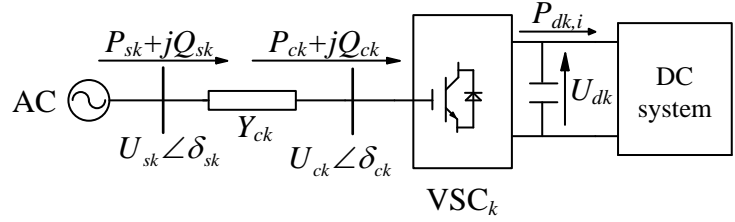

**Fig. 1 The schematic diagram of AC/DC grid**

In Fig.1, $k$ represents the $k$-th VSC connected to the DC grid, PCC (Point of Common Coupling) is the common connection point, $U_{sk}$ and $\delta_{sk}$ are the amplitude and phase angle of AC fundamental voltage at PCC, $U_{ck}$ and $\delta_{ck}$ are the fundamental voltage and phase angle at AC side of converter, $P_{sk}$ and $Q_{sk}$ are AC active power and reactive power injected at PCC point, $P_{ck}$ and $Q_{ck}$ are the active power and reactive power injected into the converter valve at the AC side of the converter station respectively; $Y_{ck}=G_{ck}+jB_{ck}$ indicates the admittance of converter transformer, $P_{d,ki}$ is the active power injected by the converter station into the DC power grid; since the filtering requirements are heavily dependent on the technology considered, the filters are not taken into account in this study. In the absence of low pass filters or with the filters omitted, the phase reactor and the converter transformer can be lumped together as described in the manuscript, eliminating the dependence on the complex filter bus voltage, the filter bus and the AC grid bus coincide, hence simplifying the equations **[20-23]**. The converter loss is converted to the AC side equivalent resistance, $r_{pu} = (P_{ac,pu} - P_{dc,pu})\left/\left(\dfrac{2}{3}P_{ac,pu}\right)^2\right.$ **[24,25]**.

The method and process of loss equivalence are detailed **in Appendix D**. So the DC system output power $P_{dk,i}$ is equal to the power injected into VSC by the AC system. The general mathematical model is shown in equation (4) **[26]**.

$$\begin{cases} P_{sk} = U_{si}^2 G_{ck} - U_{si}U_{ck}[G_{ck}\cos(\delta_{si}-\delta_{ck}) + B_{ck}\sin(\delta_{si}-\delta_{ck})] \\ Q_{sk} = -U_{si}^2 G_{ck} - U_{si}U_{ck}[G_{ck}\sin(\delta_{si}-\delta_{ck}) - B_{ck}\cos(\delta_{si}-\delta_{ck})] \\ P_{ck} = -U_{ck}^2 G_{ck} + U_{si}U_{ck}[G_{ck}\cos(\delta_{si}-\delta_{ck}) + B_{ck}\sin(\delta_{si}-\delta_{ck})] \\ Q_{ck} = U_{ck}^2 G_{ck} - U_{si}U_{ck}[G_{ck}\sin(\delta_{si}-\delta_{ck}) - B_{ck}\cos(\delta_{si}-\delta_{ck})] \\ P_{ck} = P_{dk,i} \end{cases} \tag{4}$$

When calculating AC power flow, the node voltage amplitude and phase angle are the minimum set of state variables, and the other physical quantities can be obtained by the transformation of the above variables. Similarly, for DC network, dc node voltage is the state variable of DC network. According to equation (4), the converter station is the interface of AC/DC coupling, and its related variables ($P_{ck}$, $Q_{ck}$, $U_{ck}$, $\delta_{ck}$) are directly related to the operation variables of AC node and dc node, that is, the minimum set of state variables in AC/DC power grid shall be AC node voltage (voltage amplitude $U_{si}$ and voltage phase angle $\delta_{si}$) and dc node voltage $U_{dk}$. After the converter station is equivalent to the combination of impedance and $VSC_k$ ideal devices, the equivalent model of the converter station can be obtained, as shown in Fig.2. In the figure, the active and reactive power consumed in the converter station can be expressed as equations (5).

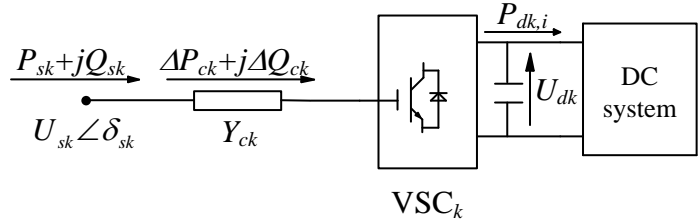

**Fig. 2 Equivalent model of converter station**

$$\begin{cases} \Delta P_{ck} = \dfrac{P_{sk}^2 + Q_{sk}^2}{U_{sk}^2} R_{ck} \\ \Delta Q_{ck} = \dfrac{P_{sk}^2 + Q_{sk}^2}{U_{sk}^2} X_{ck} \end{cases} \quad (5)$$

Since the active power injected into the AC side of $VSC_k$ is consistent with the active power injected into the dc node by $VSC_k$, it can be seen from equation (5) that the active power $P_{sk}$ injected into the converter station by the AC node can be expressed as the sum of the power $\Delta P_{ck}$ consumed by the converter station and the power $P_{dk,i}$ injected into the dc node by the converter station, i.e

$$P_{sk} = \dfrac{P_{sk}^2 + Q_{sk}^2}{U_{sk}^2} R_{ck} + P_{dk,i} \quad (6)$$

Equation (6) indirectly eliminates the relevant variables of the converter station, and directly gives the power flow mathematical model of the converter station through the state variables of AC nodes and DC nodes on both sides of the converter station.

2.3 *Control mode of converter*

At present, the control methods of multi terminal flexible DC transmission system include master-slave control, DC voltage droop control and so on. The master-slave control realizes the power balance of DC network by setting a master converter station as the constant DC voltage node, and other converter stations are set as the constant AC active power control. Reactive power control is also required for VSC converter station, including constant AC reactive power $Q_S$ and AC voltage amplitude $U_s$.

Then, there are four main types of master-slave control [17].
1) constant $P_s$ and constant $Q_s$;

2) constant $P_s$ and constant $U_s$;
3) constant $U_{dc}$ and constant $Q_s$;
4) constant $U_{dc}$ and constant $U_s$.

According to the control method of the converter, the nodes on the AC side and the DC side can be equivalent, as shown in Table 1.

As shown in Fig. 3, the voltage droop control schematic diagram is characterized by multi-point DC voltage control. If the DC voltage drops due to the power shortage in the network, the voltage drop control station increases the power injected into the network according to its own operation curve and adjusts each power output according to different $P$-$V$ operation curves. Its control characteristics can be expressed as follows:

$$\Delta f_i = (U_{dk} - U_{dkref}) + k_{droop}(P_{dk} - P_{dkref}) = 0 \tag{7}$$

Among them, $U_{dcrefi}$ and $P_{dcrefi}$ are the system operation reference nodes, $k_{droop}$ is the curve slope.

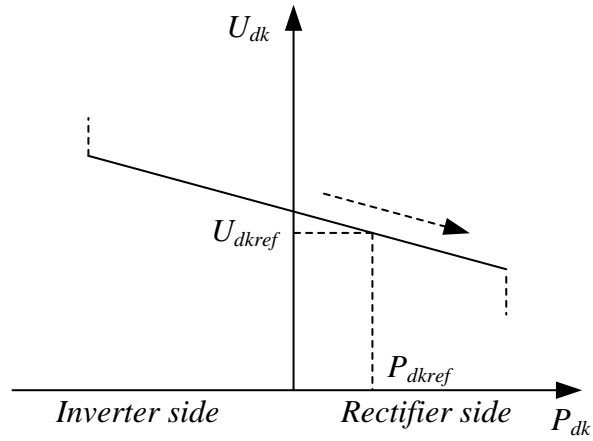

**Fig. 3 voltage-power drop control theory**

**Table 1 converter control method**

| Control mode | Control method | Control variable | AC equivalent node | DC equivalent node | classification |
|---|---|---|---|---|---|
| 1 | P-Q | active power, reactive power | $PQ$ node | Constant $P$ node | Power station |
| 2 | P-$U_s$ | active power, AC side voltage | $PV$ node | Constant $P$ node | |
| 3 | $U_{dc}$-Q | DC side voltage, reactive power | $PQ$ node | Constant $V$ node | |
| 4 | $U_{dc}$-$U_s$ | AC and DC side voltage | $PV$ node | Constant $V$ node | Voltage station |
| 5 | $U_{dc}$ droop | reactive power | $PQ$ node | Constant $V$ node | |
| 6 | $U_{dc}$ droop | AC side voltage | $PV$ node | Constant $V$ node | |

## 2.4. AC/DC power flow algorithm

### 2.4.1 Unitary system

When calculating the power flow of AC/DC power grid, the standard unit system is adopted. For ease of expression, the variable top mark "^" is used to identify the famous value. The expression of the famous value of voltage and power at VSC converter station is shown in equation (8).

$$\begin{cases} \hat{U}_{ck} = \dfrac{\sqrt{3}M_k}{2\sqrt{2}} \hat{U}_{dk} \\ \hat{P}_{dk,i} = \hat{U}_{dk} \hat{I}_{dk} \\ \hat{P}_{ck} = \dfrac{\sqrt{3}M_k}{2\sqrt{2}} \hat{U}_{dk} \hat{I}_{ck} \cos\varphi \end{cases} \quad (8)$$

Where: $M_k$ is the modulation ratio of VSC; $\varphi$ is the angle of current lag voltage at AC side of converter valve. In this paper, the loss of VSC is ignored, that is, the relationship between AC side current and DC side current of converter valve can be expressed as

$$\hat{I}_{dk} = \dfrac{\sqrt{3}M_k \cos\varphi}{2\sqrt{2}} \hat{I}_{ck} \quad (9)$$

Based on the AC power grid reference value and equation (8), the DC power grid reference value is obtained, as shown in equation (10).

$$\begin{cases} P_{dc-Base} = S_{ac-Base} \\ U_{dc-Base} = \dfrac{2\sqrt{2}}{\sqrt{3}} U_{ac-Base} \end{cases} \quad (10)$$

Therefore, the selection of current reference value and resistance reference value of DC power grid can be obtained by calculation.

Under this reference value setting, the unit value expression of AC/DC variable coupling can be obtained:

$$\begin{cases} U_{ck} = M_k U_{dk} \\ I_{ck} = \dfrac{2}{M_k \cos\varphi} I_{dk} \end{cases} \quad (11)$$

### 2.4.2 Unified iterative model of AC/DC power flow

Considering the addition of DC power grid, the power flow equation of AC node changes, which can be expressed as equation (12):

$$\begin{cases} P_{si} - P_{sk} - U_{si} \sum_{j=1}^{n_s} U_{sj}(G_{ij}\cos\delta_{ij} + B_{ij}\sin\delta_{ij}) = 0 \\ Q_{si} - Q_{sk} - U_{si} \sum_{j=1}^{n_s} U_{sj}(G_{ij}\sin\delta_{ij} - B_{ij}\cos\delta_{ij}) = 0 \end{cases} \quad (12)$$

Where: if node $i$ is the PCC, the reference values of $P_{sk}$ and $Q_{sk}$ can be obtained by reference **[27]**; otherwise, $P_{sk}$ and $Q_{sk}$ are zero.

For DC nodes, when the converter station adopts master-slave control, the state variable to be solved for determining the operating state is DC voltage $U_{dk}$, and the

corresponding power correction equation can be obtained according to formula (2) and (3):

$$P_{dk} + P_{dk,i} - U_{dk} \sum_{j=1}^{n_{dc}} G_{dkj}(U_{dk} - U_{dj}) = 0 \tag{13}$$

Where: when node $k$ is connected with converter station, $P_{dk,i}$ is equivalent to reference **[27]**; conversely, $P_{dk,i}$ is zero.

When the converter station adopts droop control, the formula (14) can be obtained for the deformation of formula (7).

$$P_{dk,i} = -1/k_{droop}\left(U_{dk,i} - U_{dkref}\right) + P_{dkref} \tag{14}$$

Therefore, when droop control is adopted, the corresponding correction equation is as follows:

$$-1/k_{droop}\left(U_{dk} - U_{dkref}\right) + P_{dkref} + P_{dk} - U_{dk}\sum_{j=1}^{n_{dc}} G_{dkj}(U_{dk} - U_{dj}) = 0 \tag{15}$$

According to Table 1, when the converter station type is voltage station, its DC side bus node is $V$ node. Since the bus voltage at $V$ node has been given as $U_{dkref}$, it is not necessary to add the correction equation of corresponding rows and columns.

If the AC/DC power grid diagram of an AC system with $n_s$ nodes and $x$ DC systems with $n_{dc}$ nodes, in which there are $m$ PV nodes in the AC power grid and 1 V nodes in the DC power grid, the number of equations that finally form the equation group is $(2n_s+n_{dc}-m-2-l)$. Combining equation (13) and equation (14) or equation (15), the matrix form of the correction equation of AC/DC power flow is obtained:

$$\begin{bmatrix} \Delta P_s - P_{sk} \\ \Delta Q_s - Q_{sk} \\ \Delta P_d + P_{dk,i} \end{bmatrix} = \begin{bmatrix} H & N+N_V & M_P \\ J & L & M_Q \\ R_\delta & R_V & X \end{bmatrix} \begin{bmatrix} \Delta \delta_s \\ \Delta U_s / U_s \\ \Delta U_d / U_d \end{bmatrix} \tag{16}$$

Where: $P_{sk}$, $Q_{sk}$ and $P_{dk,i}$ are the correction terms of AC/DC power grid coupling part to node active power and reactive power; $N_V$, $M_P$, $M_Q$, $R_\delta$ and $R_V$ are their contributions to Jacobian matrix. The latter four reflect the coupling relationship between AC and DC power grids. The expression of elements in the matrix of converter station under different control modes is shown in the reference **[27]**.

## 3. Processing method of input variables with correlation

In practice, there is correlation between photovoltaic power sources, and the premise of cumulant method is that each variable is independent of each other, so the influence of correlation should be eliminated in probabilistic power flow calculation. Specific measures are as follows:

Suppose that the correlation coefficient matrix of the input variable $Z = [z_1, z_2, \cdots, z_l]^T$ is $C_Z$:

$$C_Z = \begin{bmatrix} 1 & \rho_{z_{12}} & \cdots & \rho_{z_{1l}} \\ \rho_{z_{21}} & 1 & \cdots & \rho_{z_{2l}} \\ \vdots & \vdots & 1 & \vdots \\ \rho_{z_{l1}} & \rho_{z_{l2}} & \cdots & 1 \end{bmatrix} \tag{17}$$

Each element in the matrix can be obtained from formula (18):

$$\rho_{z_{ij}} = \rho(z_i, z_j) = \frac{C_{ov}(z_i, z_j)}{\sigma_{z_i}\sigma_{z_j}} = \frac{C_{ov}(z_j, z_i)}{\sigma_{z_i}\sigma_{z_j}} = \rho_{z_{ji}} \qquad (18)$$

According to the definition of correlation coefficient matrix, $C_Z$ is a real symmetric matrix. Therefore, the relevant sample $Z$ can be changed into an independent sample $Y$, and there is matrix $B$, so that $Y$ and $W$ meet the following relationship:

$$Y = BZ \qquad (19)$$

Where: $B$ is the conversion matrix; $Z$ is the correlation sample. According to the knowledge of probability theory, the correlation coefficient matrix between independent variables is the identity matrix e, i.e. [28]:

$$C_Y = \rho[\ BZ,\ (BZ)^T\ ] = B\rho(Z,\ Z^T)B^T = BC_ZB^T = BGG^TB^T = E \qquad (20)$$

Where: $C_Y$ is the correlation coefficient matrix of sample $Y$; $G$ is the lower triangular matrix of $C_Z$ with Cholesky decomposition; $\rho(\cdot)$ is the correlation coefficient function.

It can be seen from formula (20) that $B = G^{-1}$, which can be brought into formula (19):

$$Y = G^{-1}Z \qquad (21)$$

The uncorrelated variable $Y$ can be obtained through formula (21), and then the input variable $Z$ with correlation can be expressed as a combination of uncorrelated variables $Y$.

$$Z = GY \qquad (22)$$

## 4. Calculation of cumulants of input variables

4.1. *Probability model and cumulant calculation of photovoltaic power generation system*

Because the light intensity is affected by meteorological conditions with a certain degree of randomness, its output power is closely related to the light intensity, so the output power is also random. The sunlight intensity can be described by Beta distribution, and its probability density function is as follows [29]

$$f(p_M) = \frac{\Gamma(\alpha+\beta)}{\Gamma(\alpha)\cdot\Gamma(\beta)} \cdot (\frac{p_M}{R_M})^{\alpha-1} \cdot (1-\frac{p_M}{R_M})^{\beta-1} \qquad (23)$$

Where: $\alpha$ and $\beta$ are the shape parameters of the Beta distribution, and $\Gamma$ is the Gamma function. $R_M$ is the maximum output active power of the photoelectric, in which $R_M$ can be obtained by the following formula.

$$R_M = r_{max} \cdot \sum_{m=1}^{M} A_m \cdot \frac{\sum_{m=1}^{M} A_m \cdot \eta_m}{\sum_{m=1}^{M} A_m \cdot} \qquad (24)$$

Where: $r_{max}$ is the maximum light intensity of a certain area in a certain period.

From the sample data of the light intensity, the average $\mu$ and standard deviation $\sigma$ of the sample can be obtained, and then the shape parameters of the Beta distribution can be obtained, as shown in the following formula.

$$\begin{cases} \alpha = \mu\left[\dfrac{\mu(1-\mu)}{\sigma^2} - 1\right] \\ \beta = (1-\mu)\left[\dfrac{\mu(1-\mu)}{\sigma^2} - 1\right] \end{cases} \quad (25)$$

The *k*-th order origin moment of photovoltaic is as follows.

$$\alpha_k = \frac{\alpha(\alpha+1)\cdots(\alpha+k-1)}{(\alpha+\beta)(\alpha+\beta+1)\cdots(\alpha+\beta+k-1)} \quad (26)$$

Where: $\alpha$ and $\beta$ are the shape parameters of beta distribution. Then, the cumulants of each order of photovoltaic output can be obtained according to the following formula (27) [30].

$$\begin{cases} \gamma_1 = \alpha_1 \\ \gamma_{v+1} = \alpha_{v+1} - \sum_{j=0}^{v} C_v^j \alpha_j \gamma_{v-j+1} \end{cases} \quad (27)$$

Where: $\alpha_k$ and $\gamma_k$ represent the *k*-order origin moment and *k*-order cumulant of random variables respectively.

4.2. *Load probability distribution model and cumulant calculation*

The uncertainty of the load is mainly caused by the random fluctuation of the load. Generally, it can be described by a normal distribution, which can be expressed as [31]

$$\begin{cases} f(P_L) = \dfrac{1}{\sqrt{2\pi}\sigma_{P_L}} \exp\left(-\dfrac{(P_L - \mu_{P_L})^2}{2\sigma_{P_L}^2}\right) \\ f(Q_L) = \dfrac{1}{\sqrt{2\pi}\sigma_{Q_L}} \exp\left(-\dfrac{(Q_L - \mu_{Q_L})^2}{2\sigma_{Q_L}^2}\right) \end{cases} \quad (28)$$

Where: $\mu_{P_L}$ and $\mu_{Q_L}$ are the mathematical expectation of load active and reactive power respectively, $\sigma_{P_L}^2$ and $\sigma_{Q_L}^2$ are the variance of load active and reactive power respectively.

Load fluctuation can usually be described by formula (28). For normally distributed loads, the first-order cumulant is equal to mathematical expectation, the second-order cumulant is equal to the variance, and the third-order and above cumulants are zero, namely:

$$\begin{cases} \gamma_1 = \mu \\ \gamma_2 = \sigma^2 \\ \gamma_3 = \gamma_4 = \ldots = 0 \end{cases} \quad (29)$$

4.3. *Calculation of cumulants of input variables with unconventional distribution*

4.3.1 *Obtaining cumulants based on Monte Carlo sampling*

For the input variables that obey other distribution functions or whose distribution functions are unknown, it is difficult to obtain the analytical expression of their cumulants. Therefore, Monte Carlo sampling method can be used to calculate its cumulants.

The mathematical theory of Monte Carlo sampling is based on the law of large numbers, that is, the frequency of the event converges to the probability of the event according to the probability. As long as the sample size is large enough, the accuracy of Monte Carlo sampling can be guaranteed.

For the case where the distribution function of the input variable $W$ is known, $N$ samples $\{w_{s1}, w_{s2}, \cdots, w_{sN}\}$ can be obtained by Monte Carlo sampling technology according to its distribution function, and then the origin moment $\alpha_k$ of each order can be calculated:

$$\alpha_k = \frac{1}{N}\sum_{i=1}^{N} w_i^k, k=1,2,\cdots \tag{30}$$

Then, the cumulants of each order are obtained from the relationship formula (27) between the cumulants and the origin moment $\gamma_k$.

When the distribution function of input variable $W$ is unknown, the measured discrete historical data of input variable $W$ can be directly used as samples, and then its cumulant can be calculated according to the above method. It can be seen that the method based on Monte Carlo sampling can easily obtain the cumulants of input variables. In fact, without considering the calculation speed, this method can be used to solve the cumulants of any input variable, and is not affected by whether the distribution function of the input variable is known or not.

4.3.2 *Calculation of cumulants of input variables with correlation*

For the case that the input variables are correlated, the cumulants of the input variables are obtained based on the samples of the input variables that meet the correlation conditions. Given the marginal cumulative distribution function $F(W)$ and correlation coefficient matrix $C_W$ of the input variable $W$, the sample $W_s$ of the input variable w can be obtained through the following basic steps [32]:

1) The samples of independent standard normal distribution variable $E$ is generated by Monte Carlo sampling technology.

2) The correlation coefficient matrix $C_W$ is converted to obtain the correlation coefficient matrix $C_Q$, in which the off diagonal elements of $C_Q$ can be obtained by equation (31).

$$\rho_{q_{ij}} = D(\rho_{w_{ij}})\rho_{w_{ij}} \tag{31}$$

Where: $\rho_{q_{ij}}$ and $\rho_{w_{ij}}$ are the elements of $C_Q$ and $C_W$ in row $i$ and column $j$ respectively; the solution method is shown in literature [33].

3) The sample $\boldsymbol{Q_s}$ of the standard normal distribution variable $Q$ with the correlation coefficient matrix $C_Q$ is obtained by equation (32).

$$Q_s = G_Q E_S \tag{32}$$

Where: $G_Q$ is the lower triangular matrix obtained by Cholesky decomposition of matrix $C_Q$.

4) The sample $W_s$ of the input variable $W$ whose correlation coefficient matrix is $C_W$ can be generated by the equal probability transformation principle.

$$W_s = F^{-1}(\Phi(Q_s)) \tag{33}$$

# 5. Probabilistic Power Flow Calculation of AC/DC Hybrid System Based on cumulant method

Firstly, the deterministic power flow solution model of AC/DC hybrid system has been established in **Section 2**, and the steady-state power flow results after power system convergence can be obtained. Secondly, by decoupling the AC/DC hybrid system and linearizing it at the reference operating point of the system, the cumulants of each order of the disturbance variables are solved. Then, according to the AC/DC power flow calculation results, the corresponding sensitivity matrix of AC/DC system is obtained, so as to calculate the cumulants of each order of state variables, finally, the CDF and PDF of the output variable are obtained from the cumulant of the output variable and Gram-Charlie series.

Since the photovoltaic output power and load demand in the existing AC/DC hybrid system have great uncertainties, the probability distribution model must first be established.

### 5.1. *Linearization model of power flow equation*

The AC/DC power flow model with node voltage as state variable has been established, as shown in formula (16), and the following can be obtained by modifying formula (16)[27]:

$$\begin{bmatrix} \Delta \delta_s \\ \Delta U_s / U_s \\ \Delta U_d / U_d \end{bmatrix} = \begin{bmatrix} H & N+N_V & M_P \\ J & L & M_Q \\ R_\delta & R_V & X \end{bmatrix}^{-1} \begin{bmatrix} \Delta P_s \\ \Delta Q_s \\ \Delta P_d \end{bmatrix} + \begin{bmatrix} H & N+N_V & M_P \\ J & L & M_Q \\ R_\delta & R_V & X \end{bmatrix}^{-1} \begin{bmatrix} -P_{sk} \\ -Q_{sk} \\ P_{dk,i} \end{bmatrix} \quad (34)$$

The general form of formula (35) is:

$$\Delta X = J_0^{-1} \Delta W + J_0^{-1} P_\Delta \quad (35)$$

Therefore, the relationship between node voltage and branch power flow can be obtained:

$$\Delta H = G_0 \Delta X = G_0(J_0^{-1} \Delta W + J_0^{-1} P_\Delta) = G_0 J_0^{-1} \Delta W + G_0 J_0^{-1} P_\Delta \quad (36)$$

Where: The subscript 0 indicates the reference operating point. $G_0$ is the matrix obtained by calculating the first-order partial derivative of branch power to node voltage, and $J_0$ is the Jacobian matrix of the last iteration of AC/DC power flow.

Let $S_0 = J_0^{-1}$, $T_0 = G_0 J_0^{-1}$, $X_\Delta = J_0^{-1} P_\Delta$, $H_\Delta = G_0 J_0^{-1} P_\Delta$, the AC/DC hybrid probabilistic power flow model can be obtained:

$$\begin{cases} \Delta X = S_0 \Delta W + X_\Delta \\ \Delta H = T_0 \Delta W + H_\Delta \end{cases} \quad (37)$$

Where: $\Delta W$, $\Delta X$ and $\Delta H$ are the changes of variable injection power $W$, node voltage $X$ and branch power flow $H$ respectively, $S_0$ and $T_0$ are sensitivity matrices.

$$\begin{cases} x_i = x_{i0} + \Delta x_i = x_{i0} + \sum_r s_{ir0} \Delta w_r + x_{i\Delta} = x_{i0}' + \sum_r s_{ir0} \Delta w_r \\ h_i = h_{i0} + \Delta h_i = h_{i0} + \sum_r t_{ir0} \Delta w_r + h_{i\Delta} = h_{i0}' + \sum_r t_{ir0} \Delta w_r \end{cases} \quad (38)$$

Given the basic operation conditions of the system, the node state variable $X_0$, branch power flow variable $H_0$ and Jacobian matrix $J_0$ of the reference operation point can be obtained by the deterministic power flow calculation method, and then the sensitivity matrices $S_0$ and $T_0$ can be obtained.

### 5.2. Calculation of cumulants of state variables

It is assumed that the injection power of all nodes is independent of each other. The random disturbance $\Delta w_i$ of the injected power of node $i$ is mainly composed of the random variables of the photovoltaic injected power and load injected power of the node (i.e. $\Delta w_{Gi}$ and $\Delta w_{Li}$). The convolution operation is replaced according to the additivity of cumulants to reduce the amount of calculation. The $k$-order cumulant $\Delta w_i^k$ of the injected power of node $i$ is

$$\Delta w_i^{(k)} = \Delta w_{Gi}^{(k)} + \Delta w_{Li}^{(k)} \tag{39}$$

Where: $\Delta w_{Gi}^{(k)}$ and $\Delta w_{Li}^{(k)}$ are the $k$-order cumulants of generator injection power and load injection power of node $i$, respectively.

Since the cumulant is homogeneous, if the random variables $Y$ and $X$ satisfy $Y=aX+b$, the $k$-order cumulant relationship between $Y$ and $X$ is

$$\gamma_k(Y) = \begin{cases} a\gamma_k(X)+b, & k=1 \\ a^k\gamma_k(X), & k>1 \end{cases} \tag{40}$$

Formula (41) represents the cumulants of each order of the output variable, which is obtained by formulas (37) and (40).

$$\begin{cases} \Delta X = S_0 \Delta W + X_\Delta, \Delta H = T_0 \Delta W + H_\Delta, k=1 \\ \Delta X^{(k)} = S_0^{(k)} \Delta W^{(k)} & k>1 \\ \Delta H^{(k)} = T_0^{(k)} \Delta W^{(k)} & k>1 \end{cases} \tag{41}$$

When the injection power is correlated, equations (39) and (40) need to be corrected. Firstly, the correlation between different injection powers of the same node is analyzed. Suppose that node $i$ has $a$ injection power, of which $j(0 \leq j \leq a)$ injection powers $w_{i1}$, $w_{i2}$, ···, $w_{ij}$ are correlated. Because different injection powers of the same node are additive, the total injection power $w_{ci}$ can be obtained by summing the $j$ correlated injection powers, and the total injection power $w_{ci}$ and injection power $w_{ib}(b \neq 1,2,···,j)$ are independent of each other. Calculate the cumulants of each order of the total injection power $w_{ci}$ and other independent injection power $w_{ib}$, and modify formula (39), i.e

$$\Delta w_i^{(k)} = \Delta w_{ib}^{(k)} + \Delta w_{ci}^{(k)} \tag{42}$$

Where: $\Delta w_{ib}^{(k)}$ and $\Delta w_{ci}^{(k)}$ are the independent injection power of node $i$ and the $k$-th order cumulant of the correlated injection power, respectively.

When the injected powers $w_1, w_2, ···, w_k$ of different nodes 1, 2, ···, $k(0 \leq k \leq n)$ are correlated, and the correlation coefficient matrix is known, $k$ uncorrelated variables can $y_1, y_2, ···, y_k$ be obtained by formula (5). Then use equation (6) to express the $k$ correlated node injection power as a linear combination of uncorrelated variables $y_1, y_2, ···, y_k$:

$$w_j = \sum_{m=1}^{j} g_{jm} y_m, j=1,2,···k \tag{43}$$

Substituting formula (43) into formula (38), we can get:

$$\begin{cases} x_i = x_{i1} + \sum_r s_{ir1} \Delta w_r^{'} \\ h_i = h_{i1} + \sum_r t_{ir1} \Delta w_r^{'} \end{cases} \tag{44}$$

Where:

$$\begin{cases} s_{ir1} = s_{ir0}, t_{ir1} = t_{ir0} & r \neq 1,2,\cdots k \\ s_{ir1} = \sum_{m=r}^{k} s_{im0} g_{mr}, t_{ir1} = \sum_{m=r}^{k} t_{im0} g_{mr}, r = 1,2,\cdots k \\ x_{i1}^{'} = x_{i0}^{'}, h_{i1} = h_{i0}, \\ \Delta w_{r}^{'} = \Delta w_{r}, & r \neq 1,2,\cdots k \\ \Delta w_{r}^{'} = \Delta y_{r}, & r = 1,2,\cdots k \end{cases} \quad (45)$$

Since the input variable $W^{'} = [w_1^{'}, w_2^{'}, \cdots, w_n^{'}]^T$ is uncorrelated, the $k$-order cumulant $\Delta W^{'(k)}$ of the input variable $W^{'}$ is known, and equation (41) can be modified:

$$\begin{cases} \Delta X = S_1 \Delta W^{'} + X_{\Delta}, \Delta H = T_1 \Delta W^{'} + H_{\Delta}, k = 1 \\ \Delta X^{(k)} = S_1^{(k)} \Delta W^{'(k)} & k > 1 \\ \Delta H^{(k)} = T_1^{(k)} \Delta W^{'(k)} & k > 1 \end{cases} \quad (46)$$

According to equation (41) or equation (46), the cumulants of $X$ and $H$ can be obtained, and then the probability distribution can be obtained by the series expansion method.

5.3. *Expansion of Gram - Charlie series*

After the cumulants of each order of state variables are obtained, the probability distribution curves of node voltage and line power flow are obtained by Gram-Charlier series expansion method. Normalize the cumulant [34], let

$$g_v = \frac{\gamma_v}{\sigma^v} = \frac{\gamma_v}{\gamma^{v/2}} \quad (47)$$

Where: $g_v$ is called $v$-order normalized cumulant; $\sigma$ is the standard deviation. For any random variable $X$, assuming its expected value and standard deviation are $\mu$ and $\sigma$ respectively, its standardized form is:

$$\bar{x} = \frac{X - \mu}{\sigma} \quad (48)$$

$F(x)$ and $f(x)$ respectively represent the cumulative distribution function and probability density function of random variable $x$ and they satisfy $f(x) = F'(x)$. Therefore, $F(x)$ and $f(x)$ can be expressed as [34]:

$$F(\bar{x}) = \int_x^{\infty} \varphi(\bar{x}) dx + \varphi(\bar{x})[\frac{g_3}{3!} H_2(\bar{x}) + \frac{g_4}{4!} H_3(\bar{x}) + \frac{g_5}{5!} H_4(\bar{x}) + \frac{g_6 + 10 g_3^2}{6!} H_5(\bar{x}) + \frac{g_7 + 35 g_3 g_4}{7!} H_6(\bar{x}) + \frac{g_8 + 56 g_3 g_5 + 35 g_4^2}{8!} H_7(\bar{x}) + ...] \quad (49)$$

$$f(\bar{x}) = \varphi(\bar{x})[1 + \frac{g_3}{3!} H_3(\bar{x}) + \frac{g_4}{4!} H_4(\bar{x}) + \frac{g_5}{5!} H_5(\bar{x}) + \frac{g_6 + 10 g_3^2}{6!} H_6(\bar{x}) + \frac{g_7 + 35 g_3 g_4}{7!} H_7(\bar{x}) + \frac{g_8 + 56 g_3 g_5 + 35 g_4^2}{8!} H_8(\bar{x}) + ...] \quad (50)$$

Where: $\varphi(x)$ is the standard normal distribution density function; $H_\gamma(\bar{x})$ is the Hermite polynomial.

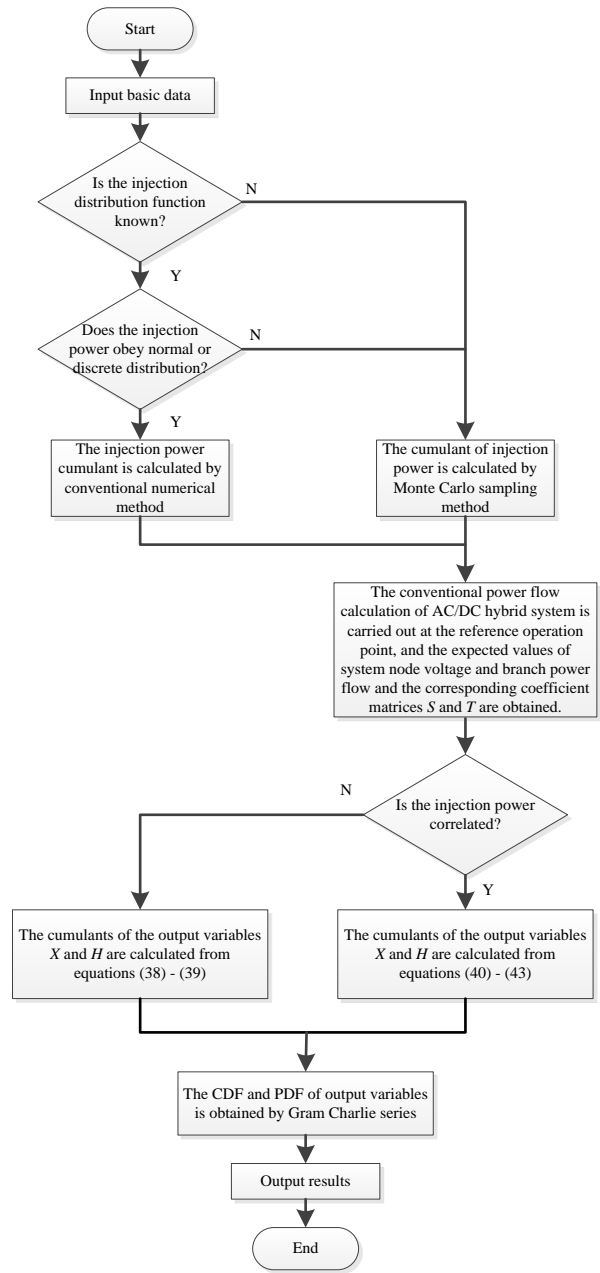

**Fig. 4 Flow chart of probabilistic power flow algorithm for AC/DC system considering the correlation of input variables**

To sum up, there are the following steps to solve the probabilistic power flow distribution of AC/DC hybrid system by using cumulant method. The flow chart of the algorithm is shown in Fig. 4.

Step 1: Input basic data (AC/DC hybrid system data, photovoltaic and load related data and distribution, control mode and the set value of each converter).

Step 2: Judge whether injection power distribution function is known and whether it obeys normal distribution or discrete distribution. If the injection power distribution function is known and obeys normal distribution or discrete distribution, the cumulant is calculated by conventional numerical method; Otherwise, the injection power cumulant is calculated by Monte Carlo sampling method.

Step 3: The conventional power flow calculation of AC/DC hybrid system is carried out at the reference operation point, and the expected values of system node voltage and branch power flow and the corresponding coefficient matrices $S$ and $T$ are obtained.

Step 4: Judge whether the injection power has correlation. If the injection powers are independent of each other, the cumulant of the output variable is obtained from equations (38) and (39). If the injection power is correlated, the cumulant of the output variable is calculated from equations (40) - (43).

Step 5: The probability density function and cumulative distribution function of state variables are obtained by Gram-Charlie series expansion method.

## 6. Case Study

### 6.1 *Case 1*

In this paper, the standard IEEE-34 bus system [35] is used, and a three-terminal DC network is added on this basis to form an AC/DC hybrid power system, as shown in Fig. 5. AC buses 14, 15 and 25 in the DC network are connected through converters. The parameters of converter, DC line parameters and converter control parameters are shown in Table 2, Table 3 and Table 4.The reference voltage of the system is $U_B$=25.64kV, the reference capacity is $S_B$=1MVA, all data formats are standard unit values, the photovoltaic module area is 2.16m$^2$, the photoelectric conversion efficiency is 13%, using constant power factor control, the maximum output active power is 0.25MW, the output reactive power is 0, and the output power obeys the Beta distribution, the shape parameters are $\alpha$=0.6799, $\beta$=1.7787. The light intensity data used in this paper are simulated by HOMER software on the monthly average value of Guangzhou, China (113°15′E, 23°7′N), and the light intensity data per hour are obtained [36]. The load data comes from literature [37], which can be reasonably assumed to be a rough approximation of normal probability density. Simulation environment is as follows: the processor is AMD Ryzen 7 4800U with Radeon Graphics 1.80GHz, the running memory is 16.0GB RAM, and the simulation software is Matlab R2018b.

Take the MCS calculation result of 10000 times as the accurate result reference. Photovoltaic is connected to node 29. Due to the fluctuation and uncertainty of photovoltaic output power, it has an impact on the power grid. To verify the effectiveness of the method, for the AC system, take the node voltage $V_{29}$ and the lines $P_{29-32}$ and $Q_{29-32}$ adjacent to node 29 as examples for analysis. For the DC system, take the voltage $V_2$ of the converter 2 and the DC line $P_{2-3}$ as examples, and obtain the probabilistic power flow calculation results through PLF-CM.

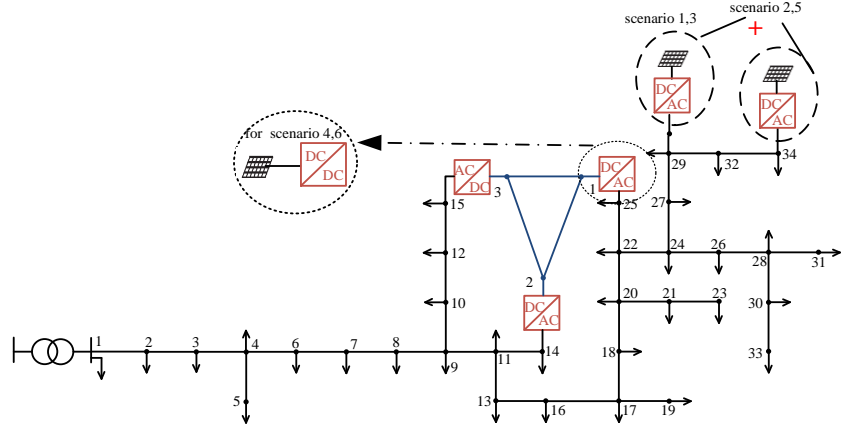

**Fig. 5 Modified IEEE 34-node test feeder system**

**Table 2 VSC parameters (p.u.)**

| VSC | Transformer parameters $Z_{tf}$ | Filter parameters $B_f$ | VSC parameters $Z_c$ |
|---|---|---|---|
| 1, 2, 3 | 0.00272+j0.203 | 0.04899 | 0.000181+j0.29743 |

**Table 3 DC power grid line parameters (p.u.)**

| Head | End | Line resistance |
|---|---|---|
| 1 | 2 | 0.0304 |
| 1 | 3 | 0.0779 |
| 2 | 3 | 0.0779 |

**Table 4 VSC control parameters（p.u.）**

| Scenario | converter | control mode | $u_{dc}$ | $U_s$ | $P_s$ | $Q_s$ | $k_{droop}$ |
|---|---|---|---|---|---|---|---|
| scenario 1,2,5 | 1 | $u$-$Q$ | 1.0 | \ | \ | 0.03 | \ |
| | 2 | $P$-$Q$ | \ | \ | 0.11 | 0.03 | \ |
| | 3 | $P$-$Q$ | \ | \ | 0.11 | 0.03 | \ |
| scenario 3 | 1 | $u$-$Q$ | 1.0 | \ | \ | 0.03 | \ |
| | 2 | $P$-$U$ | \ | 1.0 | 0.11 | \ | \ |
| | 3 | $P$-$Q$ | \ | \ | 0.11 | 0.03 | \ |
| scenario 4 | 1 | $f$-$U$ | 1.0 | \ | 1.0 | \ | \ |
| | 2 | $u$-$Q$ | 1.0 | \ | \ | 0.03 | \ |
| | 3 | $P$-$Q$ | \ | \ | 0.11 | 0.03 | \ |
| scenario 6 | 1 | $u$-$Q$ | 1.0 | \ | \ | 0 | 0.04 |
| | 2 | $P$-$U$ | \ | 1.0 | 0.11 | \ | \ |
| | 3 | $P$-$Q$ | \ | \ | 0.11 | 0 | \ |

In order to evaluate the effectiveness of the proposed method, three indexes are introduced: relative error, ARMS (Average Root Mean Square) [38] and *TIC* (Theil inequality coefficient) [39].

The relative error mainly reflects the deviation between the results obtained by this method and the reference value:

$$\varepsilon_\xi^\gamma = \left| \frac{\varepsilon_{CM}^\gamma - \varepsilon_{MCS}^\gamma}{\varepsilon_{MCS}^\gamma} \right| \times 100\% \tag{51}$$

Where: $\varepsilon_\zeta^\gamma$ is the relative error index, $\gamma$ is the output variable, and $\zeta$ is the mathematical expectation $\mu$ and standard deviation $\sigma$ of the state variable. And represent the calculated results using PLF-CM and MCS, respectively.

The ARMS indicator is used to measure the calculation accuracy of the probability distribution of the output variable.

$$\varepsilon_\gamma = \frac{\sqrt{\sum_{i=1}^{N}\left(C_{CM,i}^\gamma - C_{MCS,i}^\gamma\right)^2}}{N} \times 100\% \tag{52}$$

Where: $\varepsilon_\gamma$ is the ARMS index; $C_{CM,i}^\gamma$ and $C_{MCS,i}^\gamma$ respectively represent the value of the *i*-th point on the output variable CDF obtained by the cumulant method and the MCS method; $N$ is the number of samples on the CDF.

*TIC*, as a quantitative index, is mainly used to measure the statistical accuracy of the output variables of the proposed method; *TIC* can quantitatively evaluate the prediction accuracy of the calculation method. The value of *TIC* is between 0 and 1. The smaller the value, the closer the prediction value is to the real value, that is, the higher the prediction accuracy. The calculation formula is:

$$T^\gamma = \frac{\sqrt{\sum_{i=1}^{L}(p_{icm}^\gamma - p_{imcs}^\gamma)^2 / L}}{\sqrt{\sum_{i=1}^{L}{p_{icm}^\gamma}^2 / L} + \sqrt{\sum_{i=1}^{L}{p_{imcs}^\gamma}^2 / L}} \tag{53}$$

Where: $\gamma$ is the output variable, including node voltage amplitude, branch active power and reactive power; cm and mcs are PLF-CM and MCS respectively; $T^\gamma$ is Theil inequality coefficient; $p$ is the corresponding probability density; $L$ is the number of samples of the probability density curve.

### 6.1.1 *Simulation and method verification*

When performing AC/DC power flow calculation, according to the control method of the converter in Table 1, the PCC node can be equivalent to *PQ* or *PV* node, which can be directly applied to the existing AC/DC power flow calculation. Then, according to the location of the photovoltaic access point, the photovoltaic capacity and the different control methods of the converter, different scenarios are set for analysis.

Scenario 1, scenario 3 are connected to photovoltaic at node 29; in scenario 2, *PV* is connected to nodes 29 and 34; in scenario 4 and scenario 6, photovoltaic is connected to the AC side of converter 1 through DC transformer. The specific access location of photovoltaic modules is shown in Fig. 5.

Scenario 1: Connect photovoltaic modules at node 29, the total photovoltaic area is 4000m$^2$, the photovoltaic penetration rate is 5%, converter 1 adopts constant DC voltage control, converters 2 and 3 adopt constant active power control, and the control parameter data is shown in Table 4. The PDF and CDF of the output variables obtained using PLF-CM and MCS are shown in FA. 1.

Scenario 2: Connect photovoltaic modules at nodes 29 and 34. The total photovoltaic area is 4000m$^2$ and 800m$^2$, respectively. The converter control mode and control parameters are the same as in scenario 1. The PDF and CDF of the output variables obtained using PLF-CM and MCS are shown in FA .2.

Scenario 3: Connect photovoltaic modules at node 29, the total photovoltaic area is 4000m², the converter 2 adopts constant AC active power and AC voltage control, the control method of converter 1, 3 is the same as scenario 1, and the control parameter data is shown in Table 4. The PDF and CDF of the output variables obtained by using PLF-CM and MCS are shown in FA .3.

Scenario 4: Photovoltaic is connected to the AC side of converter 1 through DC transformer, with a total area of 4000m². Converter 1 adopts islanding control, converter 2 adopts constant DC voltage control, and converter 3 adopts constant active power control. The control parameter data are shown in Table 4. The PDF and CDF of DC system output variables obtained by PLF-CM and MCS are shown in FA .4.

Scenario 5: The influence of correlation between photovoltaic power sources on probabilistic power flow calculation results is analyzed. Photovoltaic modules are connected to nodes 29 and 34. The total photovoltaic area is 4000m² and 800m² respectively, the photovoltaic module area is 2.16m², the photoelectric conversion efficiency is 13%, and the shape parameters are $\alpha = 0.6799$, $\beta = 1.7787$. The maximum output active power is 0.25MW and the output reactive power is 0. The control mode and control parameters of each converter are the same as those in scenario 2. The correlation coefficient matrix of the two groups of photovoltaic modules is $C_{PV1}$:

$$C_{PV1} = \begin{bmatrix} 1 & 0.5 \\ 0.5 & 1 \end{bmatrix} \tag{54}$$

The probability density curve (PDF) and cumulative distribution curve (CDF) of output variables obtained by PLF-CM and MCS are shown in FA. 5.

Scenario 6: Photovoltaic is connected to the AC side of converter 1 through DC transformer, with a total area of 4000m². The droop control mode is considered in this scenario. The reference values of $U_{dcref}$, $P_{dcref}$ and $Q_{sref}$ of converter station 1 are 0.9, 0.3 and 0 respectively. The control mode and relevant data of the converter are shown in Table 4. The PDF and CDF of DC system output variables obtained by PLF-CM and MCS are shown in FA. 6.

FA. 1, FA. 2, FA. 3, FA. 4, FA. 5 and FA. 6 are in the Appendix A.

FA. 1, FA. 2, FA. 3, FA. 4, FA. 5 and FA. 6 respectively show the PDF and CDF of different output variables obtained by using MCS and PLF-CM algorithms in different scenarios. Since the photovoltaic connection is close to the photovoltaic power station on the 29 nodes, the output variable is affected by photovoltaic fluctuations larger. However, the PDF and CDF curves of the output variables obtained by the method in this paper can still maintain high calculation accuracy.

Table 5 and Table 6 show the relative error index ARMS and *TIC* index of each output variable after the three-terminal DC network is connected to the grid.

**Table 5 Relative error indicators of output variables**

| Scenario | Output variable | $\varepsilon_{\mu}^{\gamma}/10^{-2}$ | | $\varepsilon_{\sigma}^{\gamma}/10^{-2}$ | |
|---|---|---|---|---|---|
| | | $\varepsilon_{\mu.mean}^{\gamma}$ | $\varepsilon_{\mu.max}^{\gamma}$ | $\varepsilon_{\sigma.mean}^{\gamma}$ | $\varepsilon_{\sigma.max}^{\gamma}$ |
| scenario 1 | U | 0.338 | 2.350 | 1.013 | 1.856 |
| | P | 1.244 | 2.563 | 1.330 | 3.166 |
| | Q | 0.553 | 1.296 | 0.290 | 1.646 |
| scenario 2 | U | 0.294 | 1.707 | 1.165 | 2.361 |
| | P | 1.340 | 3.244 | 1.328 | 4.419 |
| | Q | 1.183 | 2.575 | 1.132 | 2.754 |

| Scenario | | | | | |
|---|---|---|---|---|---|
| scenario 3 | $U$ | 0.723 | 1.810 | 1.354 | 2.420 |
| | $P$ | 1.322 | 3.173 | 1.296 | 2.562 |
| | $Q$ | 1.545 | 3.042 | 0.449 | 1.257 |
| scenario 4 | $u_{dc}$ | 0.773 | 1.434 | 0.254 | 1.630 |
| | $P_{dc}$ | 1.464 | 2.533 | 0.725 | 1.223 |
| scenario 5 | $U$ | 0.644 | 2.206 | 1.183 | 4.405 |
| | $P$ | 1.552 | 3.739 | 0.740 | 2.553 |
| | $Q$ | 0.347 | 1.436 | 1.381 | 3.130 |
| scenario 6 | $u_{dc}$ | 0.350 | 1.584 | 0.603 | 1.440 |
| | $P_{dc}$ | 0.407 | 1.185 | 0.852 | 1.758 |

**Table 6 ARMS index and *TIC* index of output variables**

| Scenario | output variable | ARMS | | $TIC/10^{-2}$ |
|---|---|---|---|---|
| | | $\varepsilon^\gamma_{mean}/10^{-2}$ | $\varepsilon^\gamma_{max}/10^{-2}$ | |
| scenario 1 | $U$ | 0.12 | 0.47 | 1.12 |
| | $P$ | 0.32 | 0.82 | 0.31 |
| | $Q$ | 0.13 | 0.35 | 0.65 |
| scenario 2 | $U$ | 0.34 | 0.52 | 1.17 |
| | $P$ | 0.43 | 0.74 | 2.43 |
| | $Q$ | 0.25 | 0.71 | 1.63 |
| scenario 3 | $U$ | 0.18 | 0.25 | 1.42 |
| | $P$ | 0.15 | 0.43 | 0.67 |
| | $Q$ | 0.17 | 0.40 | 0.35 |
| scenario 4 | $u_{dc}$ | 0.14 | 0.33 | 1.63 |
| | $P_{dc}$ | 0.08 | 0.27 | 1.55 |
| scenario 5 | $U$ | 0.14 | 0.23 | 0.77 |
| | $P$ | 0.08 | 0.21 | 1.06 |
| | $Q$ | 0.11 | 0.43 | 0.72 |
| scenario 6 | $u_{dc}$ | 0.17 | 0.60 | 1.32 |
| | $P_{dc}$ | 0.21 | 0.56 | 0.83 |

It can be seen from Table 5 and Table 6 that for different scenarios, the accuracy of the calculation results of the method used in this paper is very small compared with MCS. The maximum values of $\varepsilon^\gamma_{\mu.mean}$, $\varepsilon^\gamma_{\mu.max}$, $\varepsilon^\gamma_{\sigma.mean}$ and $\varepsilon^\gamma_{o.max}$ are 1.552%, 3.739%, 1.381%, 4.419%, respectively. The maximum values of $\varepsilon^\gamma_{mean}$ and $\varepsilon^\gamma_{max}$ are less than 1%. *TIC* values are not more than 0.03, which is very close to 0. Analyzing *TIC* data, it can be found that PLF-CM has high prediction accuracy.

The analogy between scenario 2 and scenario 1 shows that after adding photovoltaic, the arms and tic indexes and standard deviation error of state variables increase, but the error index is still very small and still has high prediction accuracy. In scenarios 4 and 6, the effects of islanding control and droop control on the probability distribution of DC system state variables are analyzed. The average relative error of state variables is no more than 2%, and the relative error of ARMS and *TIC* is no more than 2%. Scenario 2 does not consider the impact of correlation between PV. Therefore, scenario 5 is added to consider the impact of the correlation between PV on the probabilistic power flow calculation results. It can be seen from scenario 2 and scenario 5 that the calculation results obtained after considering photovoltaic correlation are closer to MCS. At the same time, a five terminal DC network is added, and the simulation results are in Appendix C.

The simulation results show that compared with MCS method, the error of probability distribution curve obtained by the proposed method is very small. This shows that the method proposed in this paper has a certain credibility in calculating the probability power flow distribution.

### 6.1.2 Simulation time comparison

Taking scenario 3 as an example, taking the error obtained by calculating 10000 times by MCS method as the benchmark, the MCS calculation results under different sampling scales are shown in Table 7. When the sampling scale is set to 5000 and 8000 times, the maximum errors of state variables reach 5.203% and 2.462% respectively. When the sampling scale is 15000, 20000 and 50000 times, the maximum errors of state variables are not more than 1%. Under these five different sampling scales, the MCS calculation time is 20.39s, 38.85s, 100.20s ,160.74s and 798.44s respectively. Based on the above reasons, in order to more objectively reflect the calculation efficiency of the proposed algorithm, the MCS calculation time of 10000 samples is used as a reference value to evaluate the calculation efficiency of the algorithm.

**Table 7 MCS calculation error and calculation time under different sampling scales**

| Sampling scale | Output variable | $\varepsilon_{mean}^{\gamma}$ /10$^{-2}$ | Computing time/s |
|---|---|---|---|
| 5000 | $U$ | 1.413 | 20.39 |
| | $P$ | 5.203 | |
| | $Q$ | 3.161 | |
| 8000 | $U$ | 1.350 | 38.85 |
| | $P$ | 2.462 | |
| | $Q$ | 2.374 | |
| 15000 | $U$ | 0.013 | 100.20 |
| | $P$ | 0.143 | |
| | $Q$ | 0.064 | |
| 20000 | $U$ | 0.032 | 160.74 |
| | $P$ | 0.215 | |
| | $Q$ | 0.244 | |
| 50000 | $U$ | 0.041 | 798.44 |
| | $P$ | 0.242 | |
| | $Q$ | 0.565 | |

**Table 8 Comparison of calculation efficiency of different algorithms**

| Scenario | MCS/s | PLF-CM/s |
|---|---|---|
| scenario 1 | 53.03 | 0.52 |
| scenario 2 | 52.42 | 0.49 |
| scenario 3 | 56.39 | 0.42 |
| scenario 4 | 55.61 | 0.53 |
| scenario 5 | 56.47 | 1.46 |
| scenario 6 | 54.50 | 0.83 |

Table 8 shows the time taken by using two methods in different scenarios. Because scenario 1, scenario 2 and scenario 3 are affected by the converter control mode, the DC side is not affected by the random disturbance on the AC side. Therefore, there is no probability problem for the state variables in the DC system, and the calculation time is very short. In scenario 5, the correlation between PV is considered, which increases the complexity of the program. Although the calculation time is increased compared with scenario 2 (correlation is not considered), it does not exceed 2s. In scenarios 4 and 6, a converter station adopts Island control or droop control, so that the stochastic impact on the outputs of DC grid can be observed. Due to the influence of converter control mode and program complexity, the calculation time of PLF-CM increases, but the calculation time of PLF-CM is still less than 1 second, indicating that the algorithm is less affected by converter control mode, correlation and the number of photovoltaic power stations.

### *6.2 Case 2*

To verify the universality of the method, a five terminal DC power grid is connected to the IEEE57 node system, as shown in Fig. 6. Various parameters of the converter, DC line parameters and converter control parameters are shown in Table 9, Table 10 and Table 11. In addition, 6 and 7 are pure DC nodes, and the DC bus 6 is connected with 0.1p.u. DC load and 0.3p.u. DC photovoltaic power supply. Photovoltaic modules are used in the system, with an area of 2.16m$^2$, photoelectric conversion efficiency of 13%, constant power factor control, maximum output active power of 0.25MW, output reactive power of 0, output power subject to beta distribution, and shape parameters of $\alpha = 0.6799$, $\beta=1.7787$. The benchmark capacity of AC/DC hybrid power grid is 100MVA, and all data forms are standard unit values. Assuming that the load of each node follows normal distribution, the expected value is the peak load value of the original data, and the standard deviation is 20% of the expected value.

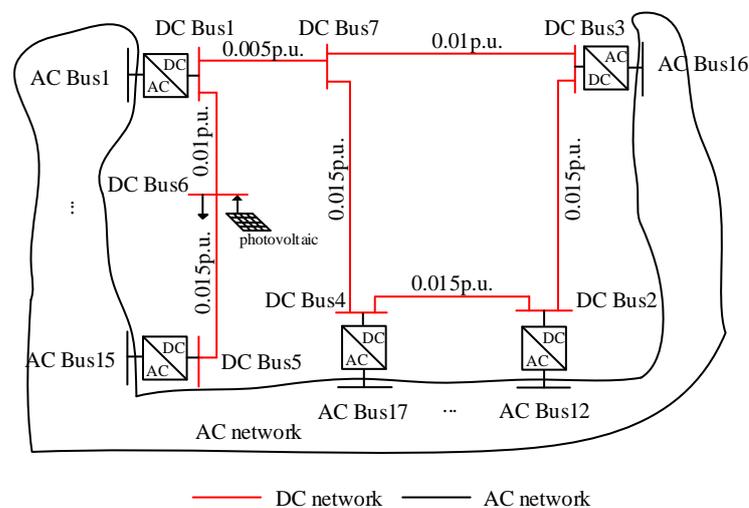

**Fig. 6 Modified IEEE57 node system**

**Table 9 VSC parameters (p.u.)**

| VSC | Transformer parameters $Z_{tf}$ | Filter parameters $B_f$ | VSC parameters $Z_c$ |
|---|---|---|---|
| 1 | 0.001+j0.033 | 0 | 0.05 |
| 2 | 0.0015+j0.05 | 0 | 0.15 |
| 3 | 0.003+j0.1 | 0.1 | 0.15 |
| 4 | 0.003+j0.1 | 0 | 0.15 |
| 5 | 0.003+j0.1 | 0 | 0.075 |

**Table 10 DC power grid line parameters (p.u.)**

| Head | End | Line resistance |
|---|---|---|
| 1 | 6 | 0.01 |

| | 1 | 7 | 0.005 |
|---|---|---|---|
| | 2 | 3 | 0.015 |
| | 2 | 4 | 0.015 |
| | 3 | 7 | 0.01 |
| | 4 | 7 | 0.015 |
| | 5 | 6 | 0.015 |

**Table 11 VSC control parameters（p.u.）**

| Scenarios | Converter | Control mode | $u_{dc}$ | $U_s$ | $P_s$ | $Q_s$ | $k_{droop}$ |
|---|---|---|---|---|---|---|---|
| scenaρio 1,2,5 | 1 | u-Q | 1.0 | \ | \ | 0.03 | \ |
|  | 2 | P-Q | \ | \ | 0.05 | 0.03 | \ |
|  | 3 | P-Q | \ | \ | -0.20 | 0.03 | \ |
|  | 4 | P-Q | \ | \ | -0.16 | 0.03 | \ |
|  | 5 | P-Q | \ | \ | 0.07 | 0.03 | \ |
| scenario 3 | 1 | u-Q | 1.0 | \ | \ | 0.03 | \ |
|  | 2 | P-U | \ | 1.0 | 0.05 | \ | \ |
|  | 3 | P-Q | \ | \ | -0.20 | 0.03 | \ |
|  | 4 | P-Q | \ | \ | -0.16 | 0.03 | \ |
|  | 5 | P-Q | \ | \ | 0.07 | 0.03 | \ |
| scenario 4 | 1 | u-Q | 1.0 | \ | -1.5 | 0 | 0.04 |
|  | 2 | u-Q | 1.0 | \ | 0.5 | 0 | 0.04 |
|  | 3 | P-Q | \ | \ | 0.4 | 0.1 | \ |
|  | 4 | P-U | \ | 1.02 | 0.6 | \ | \ |
|  | 5 | P-Q | \ | \ | 0.5 | 0 | \ |

### 6.2.1 *Simulation and method verification*

*PV* access point: In the AC system: Scenario 1 and Scenario 3 are connected to the PV at node 29; in scenarios 2 and 5, the PV is connected to nodes 29and 34; in the DC system: in scenario 4, the DC photovoltaic is connected to the DC bus 6. The specific access location of photovoltaic module is shown in Fig. 6.

Scenario 1: Connect photovoltaic modules at node 29, the total photovoltaic area is 4000m$^2$, the photovoltaic penetration rate is 5%. The control mode and relevant data of the converter are shown in Table 11. The PDF and CDF of the output variables obtained using PLF-CM and MCS are shown in FB. 1.

Scenario 2: Connect photovoltaic modules at nodes 29 and 34. The total photovoltaic area is 4000m$^2$ and 800m$^2$, respectively. The converter control mode and control parameters are the same as in scenario 1. The PDF and CDF of the output variables obtained using PLF-CM and MCS are shown in FB. 2.

Scenario 3: Connect photovoltaic modules at node 29, the total photovoltaic area is 4000m$^2$. The control mode and relevant data of the converter are shown in Table 11. The PDF and CDF of the output variables obtained by using PLF-CM and MCS are shown in FB. 3.

Scenario 4: In this scenario, droop control mode is adopted, and the control mode and reference value of converter station are shown in Table 12. The control mode and corresponding data of other converter stations are shown in Table 11. The PV is connected to DC bus 6. The PDF and CDF of DC system output variables obtained by PLF-CM and MCS are shown in FB. 4.

**Table 12 Control and references of VSC**

| Converter | Control mode | $U_{dcref}$ | $P_{dcref}$ | $Q_{sref}$ | ρ |
|---|---|---|---|---|---|
| 1 | u-Q | 0.995 | 0.30 | 0 | 0.04 |
| 2 | u-Q | 1.005 | 0.95 | 0 | 0.04 |

Scenario 5: In this scenario, considering the influence of the correlation between photovoltaic, photovoltaic modules are connected at 29 nodes and 34 nodes. The total photovoltaic area is 4000m² and 800m² respectively, the photovoltaic module area is 2.16m², the photoelectric conversion efficiency is 13%, and the shape parameter is α = 0.6799，β = 1.7787. The maximum output active power is 0.25MW and the output reactive power is 0. The control mode and control parameters of each converter are the same as those in scenario 2. The control mode and control parameters of each converter are shown in Table 11. The correlation coefficient matrix of the two groups of photovoltaic modules is $C_{PV2}$ :

$$C_{PV2} = \begin{bmatrix} 1 & 0.5 \\ 0.5 & 1 \end{bmatrix} \tag{55}$$

The probability density curve (PDF) and cumulative distribution curve (CDF) of output variables obtained by PLF-CM and MCS are shown in FB. 5.

FB. 1, FB. 2, FB.3, FB. 4 and FB. 5 are in the Appendix B.

FB. 1, FB. 2, FB.3, FB. 4 and FB. 5 respectively show PDF and CDF of different output variables obtained using MCS and PLF-CM algorithms in different scenarios, respectively. The PDF and CDF curves of output variables obtained by this method have high calculation accuracy.

Table 13 shows the relative error index of the output variable, and Table 14 shows the ARMS index and *TIC* index of the output variable.

**Table 13 Relative error indicators of output variables**

| Scenario | Output variable | $\varepsilon_\mu^\gamma/10^{-2}$ | | $\varepsilon_\sigma^\gamma/10^{-2}$ | |
|---|---|---|---|---|---|
| | | $\varepsilon_{\mu.mean}^\gamma$ | $\varepsilon_{\mu.max}^\gamma$ | $\varepsilon_{\sigma.mean}^\gamma$ | $\varepsilon_{\sigma.max}^\gamma$ |
| scenario 1 | U | 0.212 | 1.044 | 0.345 | 3.430 |
| | P | 1.064 | 3.237 | 0.352 | 3.260 |
| | Q | 0.257 | 1.074 | 0.208 | 1.620 |
| scenario 2 | U | 0.492 | 1.574 | 0.364 | 3.147 |
| | P | 1.363 | 3.832 | 1.301 | 4.767 |
| | Q | 1.535 | 3.471 | 0.405 | 2.660 |
| scenario 3 | U | 0.645 | 1.133 | 0.680 | 1.391 |
| | P | 1.197 | 2.914 | 1.323 | 5.440 |
| | Q | 1.012 | 2.107 | 0.644 | 1.143 |
| scenario 4 | $u_{dc}$ | 0.742 | 1.326 | 1.353 | 2.880 |
| | $P_{dc}$ | 0.834 | 2.720 | 1.322 | 3.457 |
| scenario 5 | U | 0.763 | 1.121 | 0.672 | 3.781 |
| | P | 1.335 | 3.930 | 0.910 | 3.735 |
| | Q | 0.740 | 3.354 | 0.632 | 2.420 |

Table 14 ARMS and *TIC* indicators of output variables

| Scenarios | output variable | ARMS | | *TIC*/$10^{-2}$ |
|---|---|---|---|---|
| | | $\varepsilon^\gamma_{mean}/10^{-2}$ | $\varepsilon^\gamma_{max}/10^{-2}$ | |
| scenario 1 | $U$ | 0.12 | 1.43 | 3.74 |
| | $P$ | 0.30 | 1.83 | 4.91 |
| | $Q$ | 0.27 | 1.75 | 2.68 |
| scenario 2 | $U$ | 0.14 | 1.32 | 3.62 |
| | $P$ | 0.29 | 1.40 | 4.22 |
| | $Q$ | 0.21 | 1.06 | 2.12 |
| scenario 3 | $U$ | 0.15 | 1.43 | 2.15 |
| | $P$ | 0.14 | 1.12 | 3.03 |
| | $Q$ | 0.14 | 1.27 | 1.42 |
| scenario 4 | $u_{dc}$ | 0.26 | 1.57 | 2.13 |
| | $P_{dc}$ | 0.28 | 1.22 | 2.34 |
| scenario 5 | $U$ | 0.14 | 1.25 | 1.40 |
| | $P$ | 0.16 | 1.16 | 2.73 |
| | $Q$ | 0.14 | 0.79 | 1.21 |

It can be seen from Table 13 and Table 14 that for different scenarios, the maximum values of $\varepsilon^\gamma_{\mu.mean}$, $\varepsilon^\gamma_{\mu.max}$, $\varepsilon^\gamma_{\sigma.mean}$ and $\varepsilon^\gamma_{\sigma.max}$ are 1.535%, 3.390%, 1.353%, and 5.440%, respectively, and the maximum values of $\varepsilon^\gamma_{mean}$ and $\varepsilon^\gamma_{max}$ are 0.30% and 1.83%, respectively. *TIC* is not more than 0.05, which is very close to 0, indicating that this method has good prediction accuracy in case 2.

The analogy between scenario 2 and scenario 1 shows that after adding photovoltaic, the mean error and standard deviation error of state variables increase, but the error indicators are very small. The maximum values of $\varepsilon^\gamma_\xi$, ARMS and *TIC* are 4.767%, 1.40% and 4.22% respectively. In scenario 4, the influence of droop control on the probability distribution of DC system state variables is analyzed, in which the maximum errors of $\varepsilon^\gamma_\xi$, ARMS and *TIC* indexes are 3.457%, 1.57% and 2.34% respectively. Scenario 2 does not consider the impact of the correlation between PV. Therefore, in scenario 5, the influence of the correlation between PVs on the probabilistic power flow calculation results is considered. It can be seen from Table 13 and Table 14 that the error index of state variables becomes smaller after considering photovoltaic correlation. After considering the correlation, the calculation accuracy of the average expected values of *U*, *P* and *Q* is improved by 0, 0.13% and 0.07% respectively. the *TIC* indexes of *U*, *P* and *Q* are less than those in scenario 2.

### 6.3 *Results for different PV generation correlations*

To evaluate the impact of PV correlation on power system uncertainty, an examination has been performed. Taking scenario 5 of case 1 as an example, it is assumed that the correlation coefficient between two photovoltaic power stations is $\rho$, because there is usually a positive correlation between adjacent *PV* [14]. When the correlation coefficient $\rho$ is set to 0.2, 0.5 and 0.8 respectively, the influence of photovoltaic correlation on system node voltage and branch power flow is analyzed.

Fig. 7 shows the CDF curve of voltage amplitude of 29 node under different correlation coefficients. Table 15 compares the mean value, standard deviation, overvoltage probability (OVP) and low voltage probability (LVP) of steady-state voltage under different correlation intensities. It can be seen that the expected value of node voltage is little affected by photovoltaic correlation, and the standard deviation is positively correlated with photovoltaic correlation coefficient. When the PV correlation coefficient increases from 0.2 to 0.8, the expected value of voltage amplitude of node 29 increases by 0.3%, and the standard deviation increases by 36.4%. The enhancement of photovoltaic correlation also increases the probability of node voltage running in low voltage section and high voltage section.

Fig. 8 and Table 16 show the probability distribution of branch power flow under different correlation strength.

To sum up, with the increase of correlation coefficient, the average value of bus voltage and branch power flow is basically unchanged, but the standard deviation increases. Therefore, the positive correlation between photovoltaic power stations will make the uncertainty problem more serious.

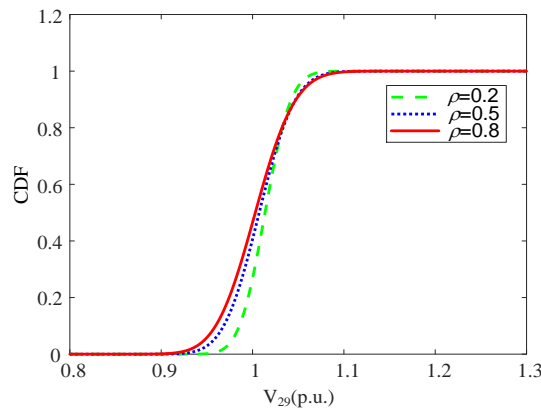

Fig. 7 voltage amplitude CDF of node 29 corresponding to different correlation coefficients

Table 15 probability distribution of node voltage under different correlation coefficients

| $\rho$ | 0.2 | 0.5 | 0.8 |
| --- | --- | --- | --- |
| $\mu$ | 0.987 | 0.990 | 0.990 |
| $\sigma$ | 0.033 | 0.040 | 0.045 |
| OVP(>1.05) | 5.35% | 8.65% | 8.83% |
| LVP(>1.1) | 0.01% | 0.17% | 0.30% |
| LVP(<0.9) | 0 | 0.02% | 0.11% |

Note: $\rho$ represents the correlation coefficient, $\mu$ and $\sigma$ represent the mean and standard deviation of state variables respectively.

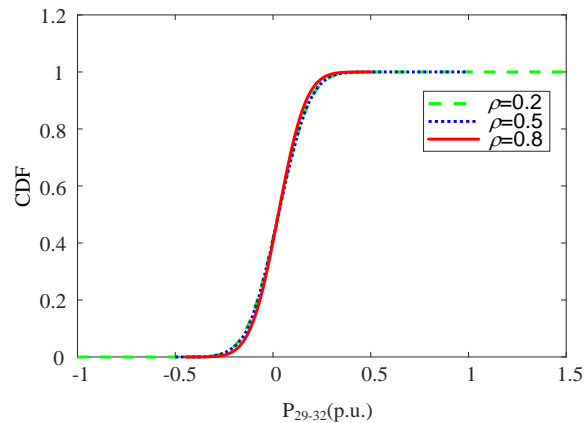

**Fig. 8 CDF of branch power flow corresponding to different correlation coefficients**

**Table 16 probability distribution of branch flow under different correlation coefficients**

| $\rho$ | 0.2 | 0.5 | 0.8 |
|---|---|---|---|
| $\mu$ | 0.027 | 0.027 | 0.027 |
| $\sigma$ | 0.030 | 0.038 | 0.048 |

### 6.4 *Sensitivity analysis of the algorithm*

At the same time, to verify the sensitivity of this algorithm to the scale of photovoltaic grid connection. Take scenario 1 in case 1 as an example, test the ARMS error of the system output variables for different photovoltaic penetration rates, and compare it with the calculation result of MCS. The result is shown in Fig. 9.

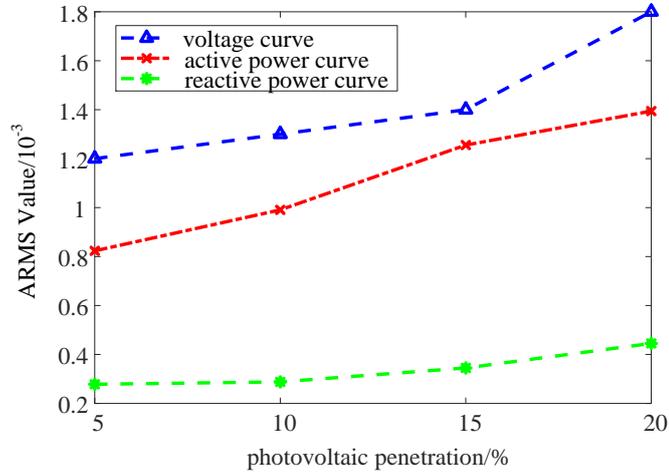

**Fig. 9 ARMS curves of output variables under different photovoltaic penetration**

As can be seen from Fig. 9, with the increase of photovoltaic penetration rate, the arms curves of node voltage and branch power obtained by the proposed algorithm increase slightly, but the error is still no more than 1%. The above results show that the algorithm proposed in this paper is suitable for stochastic power flow analysis of large-scale photovoltaic grid connected power system.

## 7. Conclusions

This paper proposes a probabilistic power flow calculation method based on the cumulant method for AC/DC hybrid systems. Simulation tests have been carried on the modified IEEE 34-bus and 57-bus systems. Based on the above analysis, the following conclusions can be drawn:

1）The power system will be affected by many uncertain factors such as photovoltaic and load. The calculation results of case 1 and case 2 verify the effectiveness of the proposed method. Compared with MCS method, it greatly shortens the calculation time, so as to win valuable time for operation dispatchers to take corresponding measures.

2）With the rapid development of DC transmission technology, DC load is also increasing. In this paper, different control modes are set for the converter, the PCC node of the converter station is equivalent, and the AC/DC probabilistic power flow calculation model is established, which can accurately obtain the distribution of the line probabilistic power flow and has good flexibility.

3）The algorithm is suitable for different VSC control modes. This algorithm can be used whether DC power grid adopts single point voltage control or multi-point voltage control. Compared with MCS method, the PDF and CDF curves of the output variables obtained by the proposed method have high fitting degree, and the ARMS error index, relative error index and *TIC* index are very small, indicating that the method used in this paper has high prediction accuracy.

4) The positive correlation between photovoltaic makes the uncertainty of power system more serious. In the correlation analysis, it can be found that with the increase of correlation coefficient, the mean value of state variables is basically unchanged, and the standard deviation of state variables increases gradually.

5）With the increase of PV permeability, the ARMS error curve of output variable increases slightly, but the ARMS error value of output variable is very small. The results show that this method maintains high calculation accuracy in probabilistic power flow analysis, and has strong adaptability and robustness to the scale of photovoltaic grid connection.

## Acknowledgment

This work was supported by the National Natural Science Foundation of China(U2066208), national Key R&D program of China（No.2018YFB0904600），Key Project of Jilin Science and Technology Bureau（2019301163）.

## Biographies

**Yinfeng Sun** was born in Siping, Jilin province, China. He received his B.Sc. degree and M.Sc degree in Electrical Engineering from Northeast Electric Power University, and Ph.D. degree from North China Electric Power University in 2017. His employment experience included the experimentalist of Northeast Electric Power University, Jilin, China, from 2009 until now. He is currently a associate professor in the Department of Electrical Engineering, Northeast Electric Power University, China. His research interests are in the areas of power system operation and optimization, stability and security assessment, power system planning and Flexible HVDC system modeling and simulation analysis.

**Dapeng Xia** received his bachelor's degree in electrical engineering from Luoyang Institute of Science and Technology in 2015. At present, he is also studying for a master's degree in Northeast Electric Power University. His research interests are in the areas of Probabilistic flow calculation.

**Zichun Gao** received a bachelor's degree in electrical engineering from Northeast Electric Power University in 2020. He is pursuing a master's degree still in Northeast Electric Power University now. His research interests are in the areas of Probabilistic flow calculation.

**Zhenhao Wang** was born in Weifang, Shandong Province, China. The research direction is to monitor and diagnose the operating status of power transmission and transformation equipment, energy saving and optimization of operation of distribution systems, and grid control of new energy sources.

**Guoqing Li** received his B.Sc.degree and M.Sc degree in Electrical Engineering from Northeast Electric Power University, and Ph.D. degree from Tianjin University in 1998. He is currently a professor in the Department of Electrical Engineering, Northeast Electric Power University, China. His research interests are in the areas of power system operation and optimization, stability and security assessment and power system planning.

**Weihua Lu** received a bachelor's degree in electrical engineering from Northeast Electric Power University in 2019. He is pursuing a master's degree still in Northeast Electric Power University now. His research interests are in the areas of Probabilistic flow calculation.

**Xueguang Wu** received his B.Sc. degree in Electrical Engineering from Northeast Dianli University in 1988 and M.Sc degree from China Electric Power Research Institute in 1996 and Ph.D. degree from Wuhan University in 2000. He is currently national distinguished Expert of "1000 elite grogram" in Smart Grid Research Institute, SGCC and Professor in the Department of Electrical Engineering, Northeast Dianli University, China. His research interests are in the areas of HVDC control amd modelling, renewable energy integration, power system operation and stability.

**Yang Li** received his Ph.D. degree in Electrical Engineering from North China Electric Power University (NCEPU), Beijing, China, in 2014. He is a professor at the School of Electrical Engineering, Northeast Electric Power University, Jilin, China. From Jan. 2017 to Feb. 2019, he was also a postdoc with Argonne National Laboratory, Lemont, United States. His research interests include power system stability and control, renewable energy integration, energy storage and smart grids. He is a senior member of the IEEE, and serves as an Associate Editor for the journals of IET Renewable Power Generation, IEEE ACCESS, and Scientific Reports.

# Appendix A

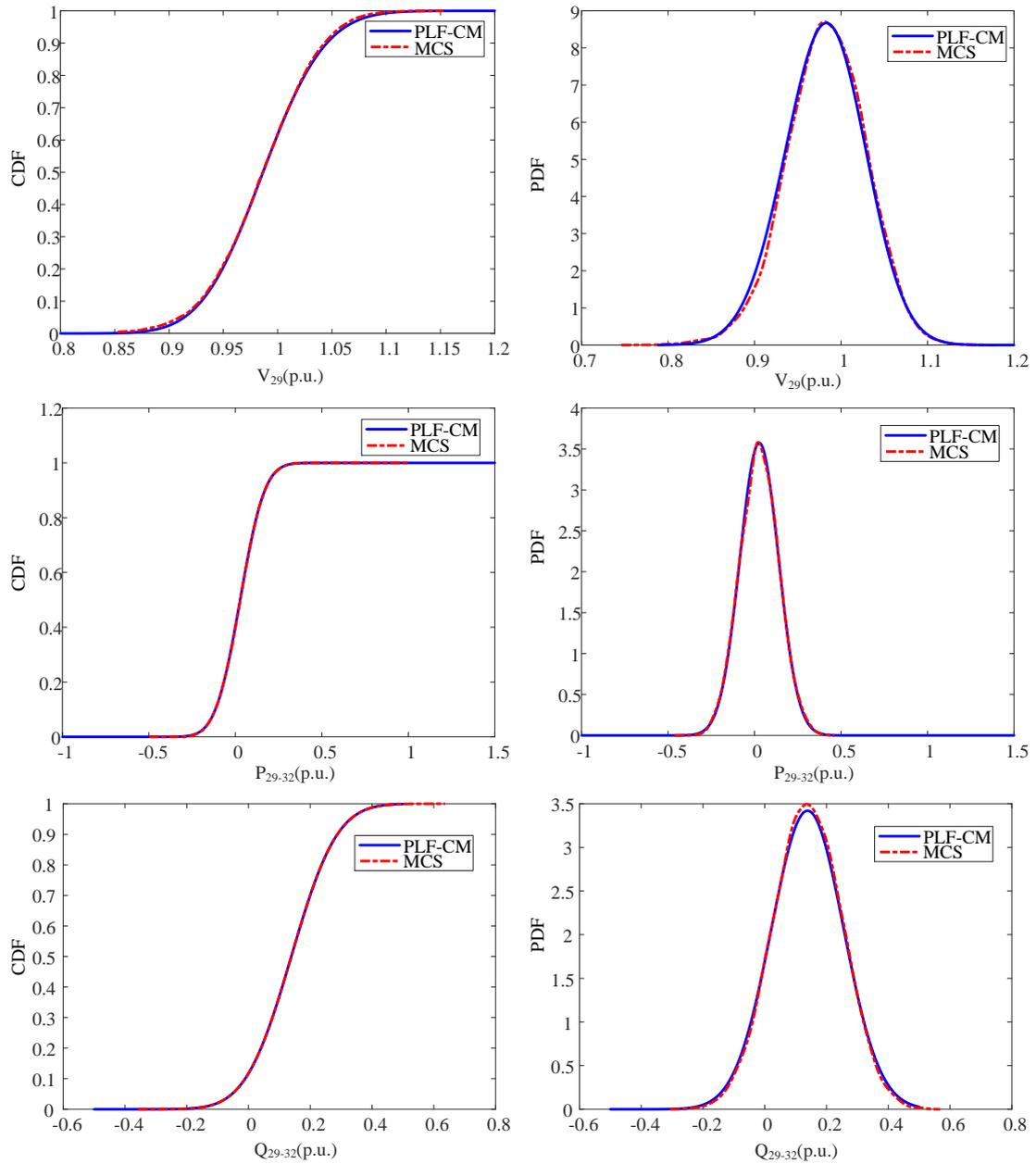

**FA. 1 PDF and CDF of output variables**

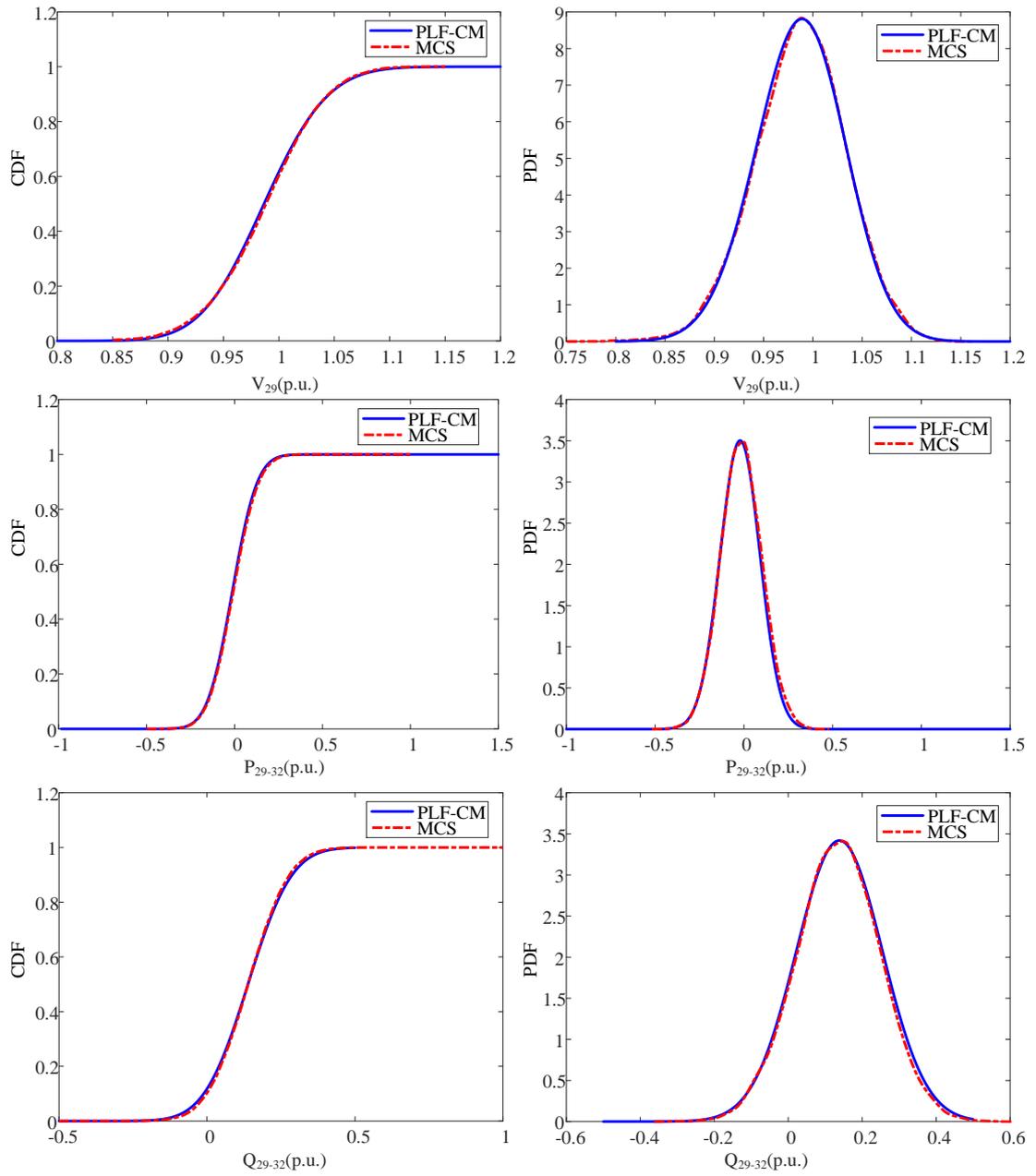

FA. 2 PDF and CDF of output variables

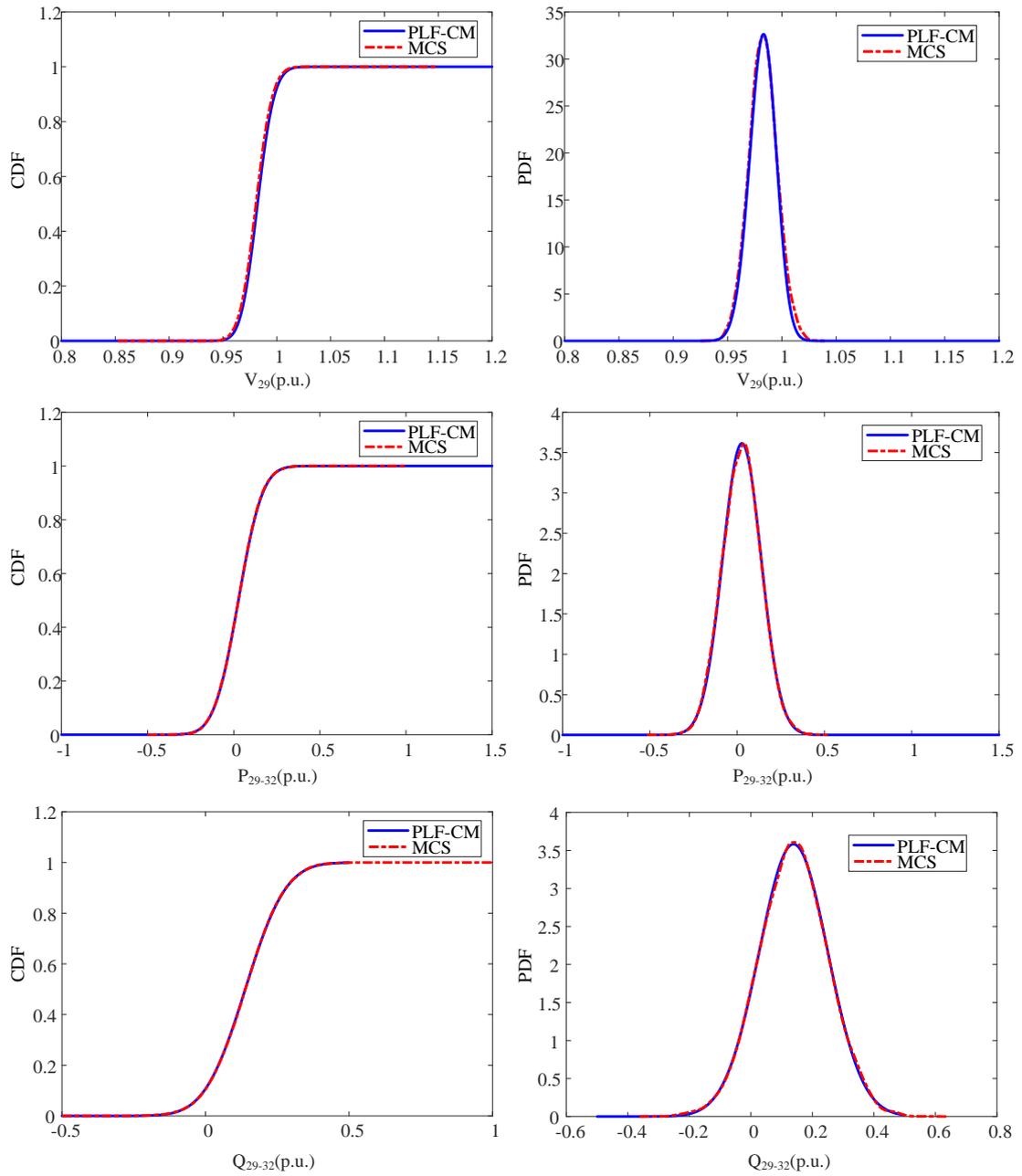

**FA. 3 PDF and CDF of output variables**

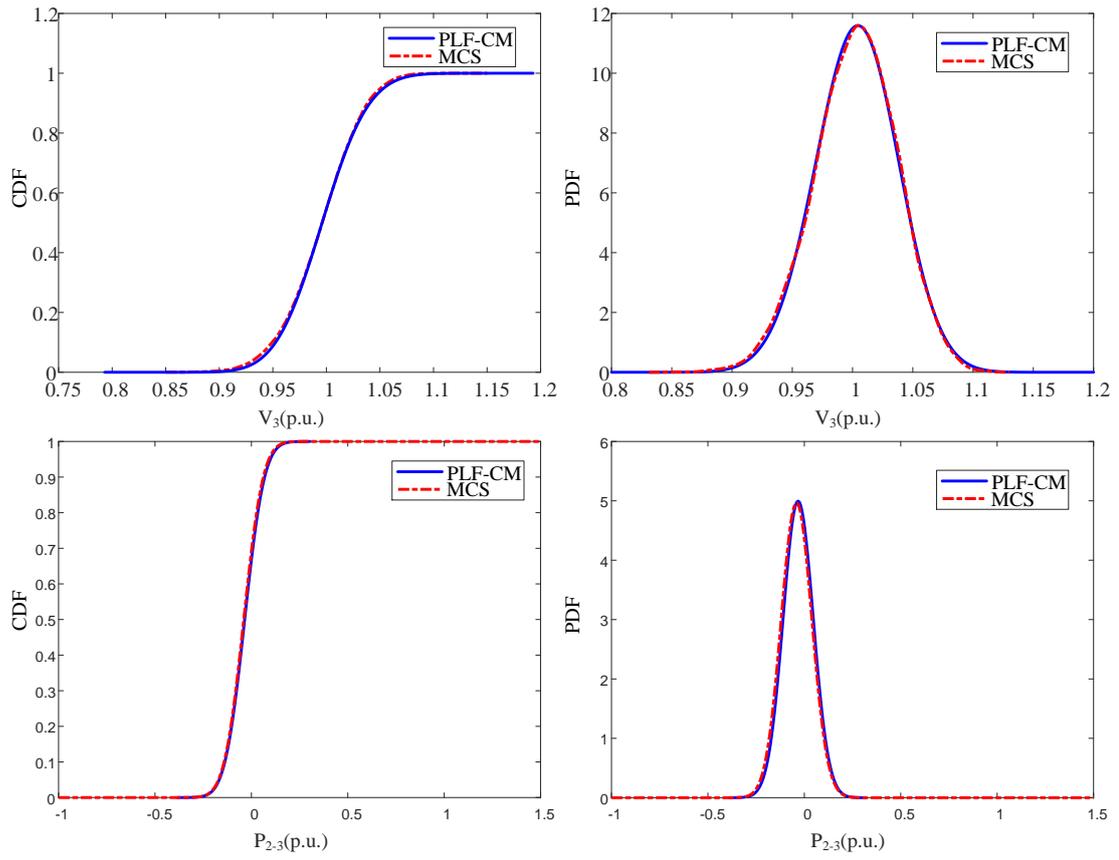

**FA. 4 PDF and CDF of output variables**

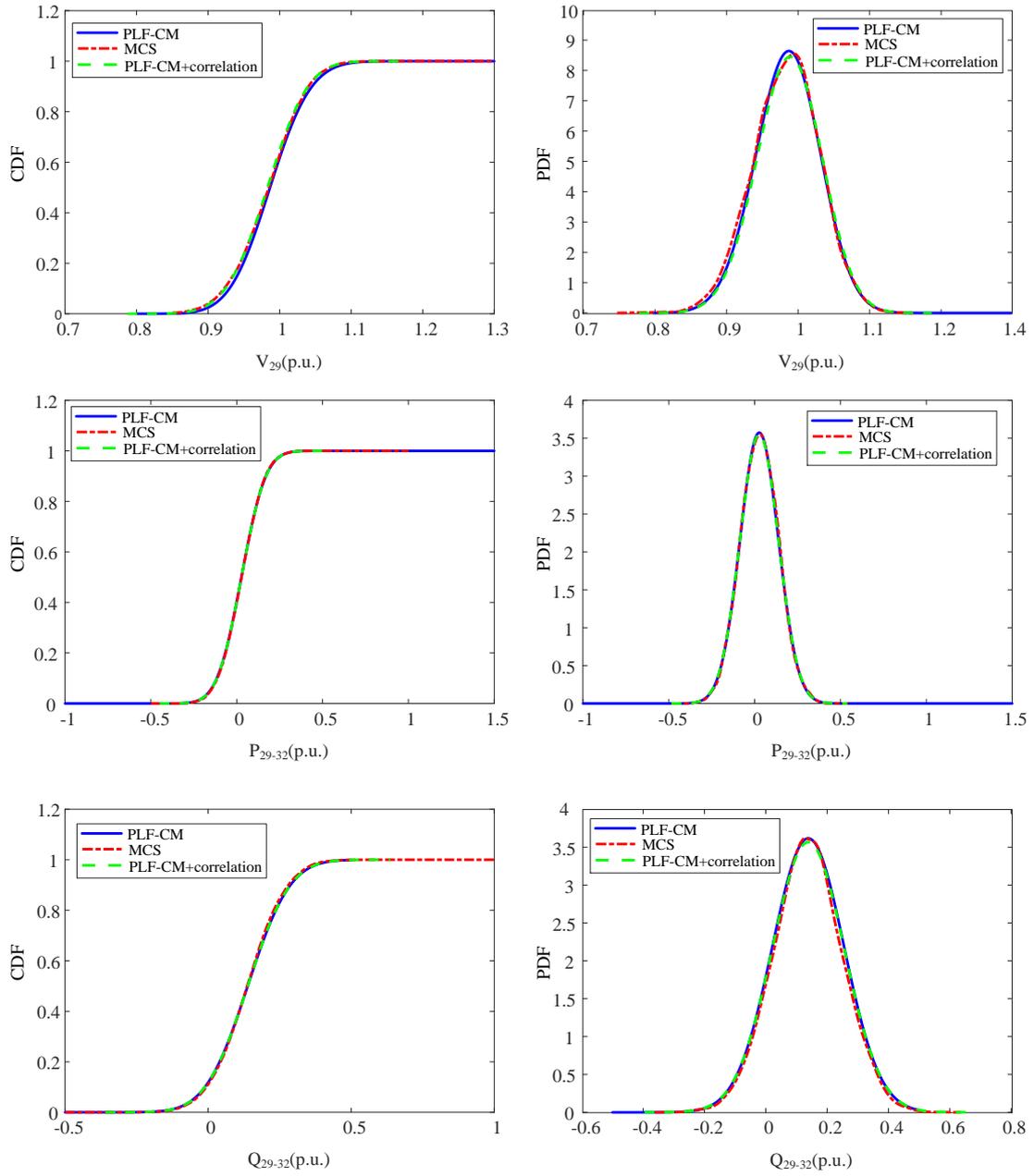

**FA. 5 PDF and CDF of output variables**

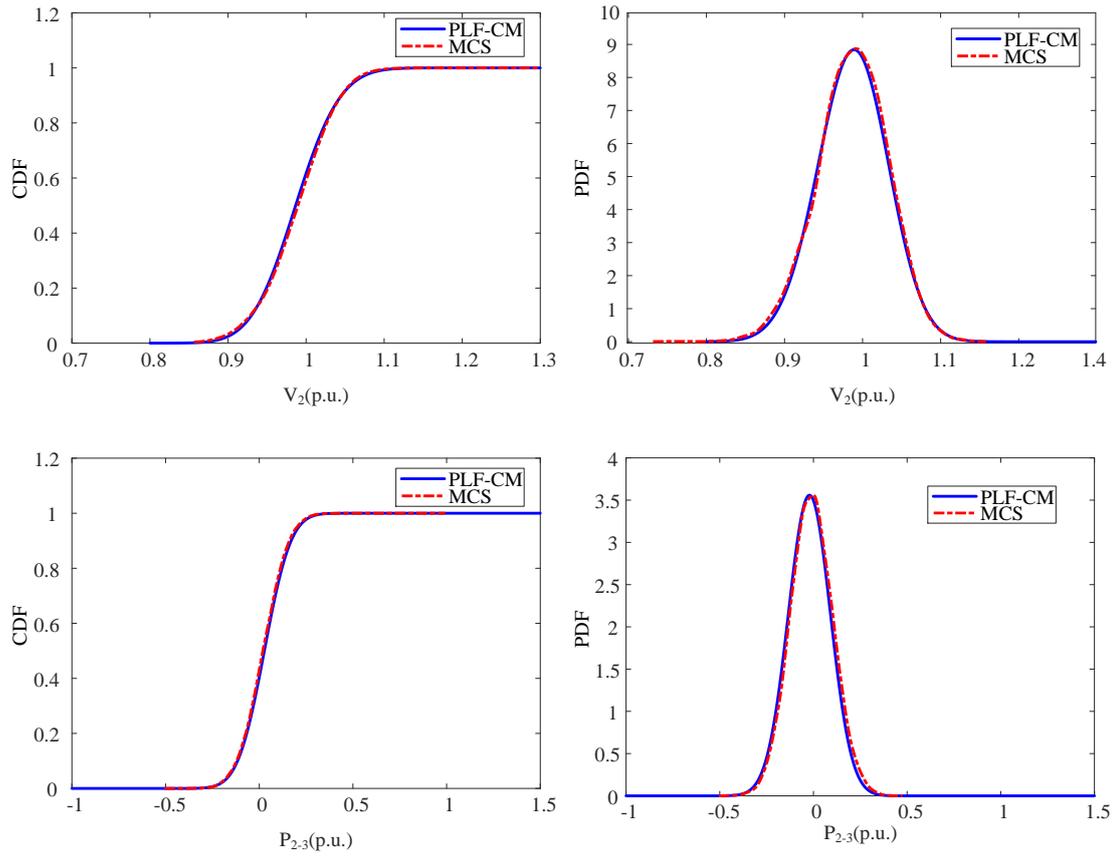

**FA. 6 PDF and CDF of output variables**

# Appendix B

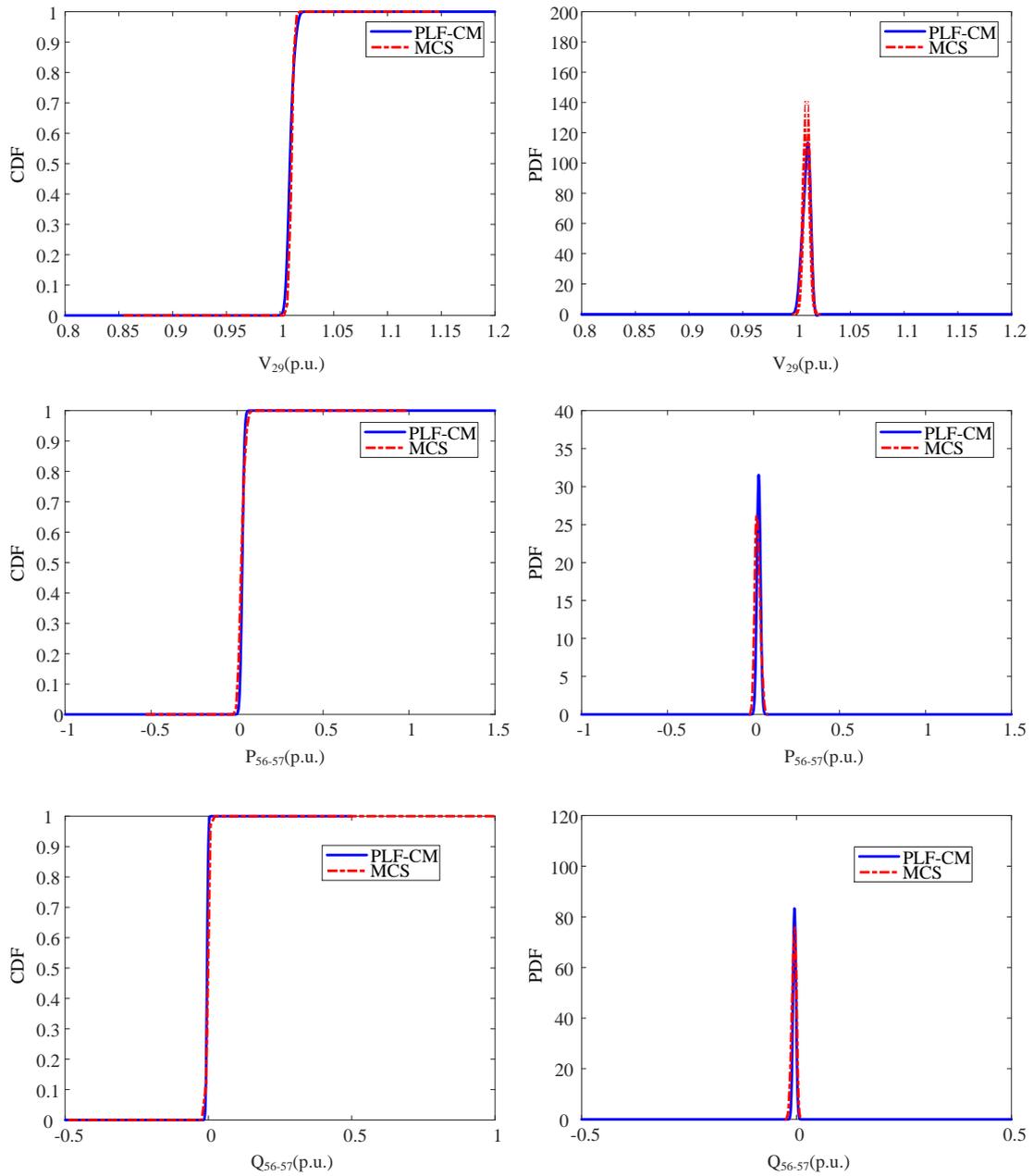

**FB. 1 PDF and CDF of output variables**

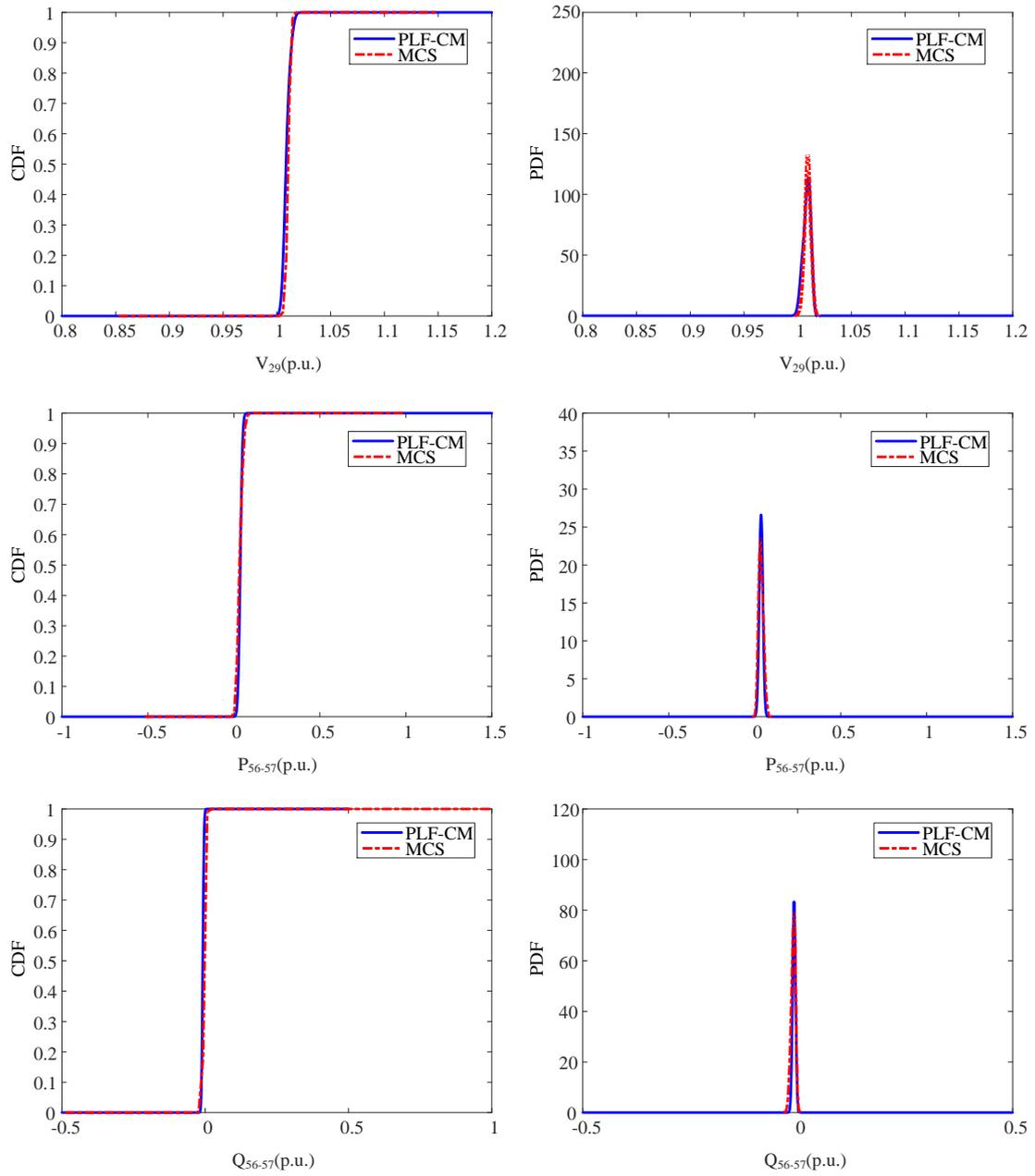

**FB. 2 PDF and CDF of output variables**

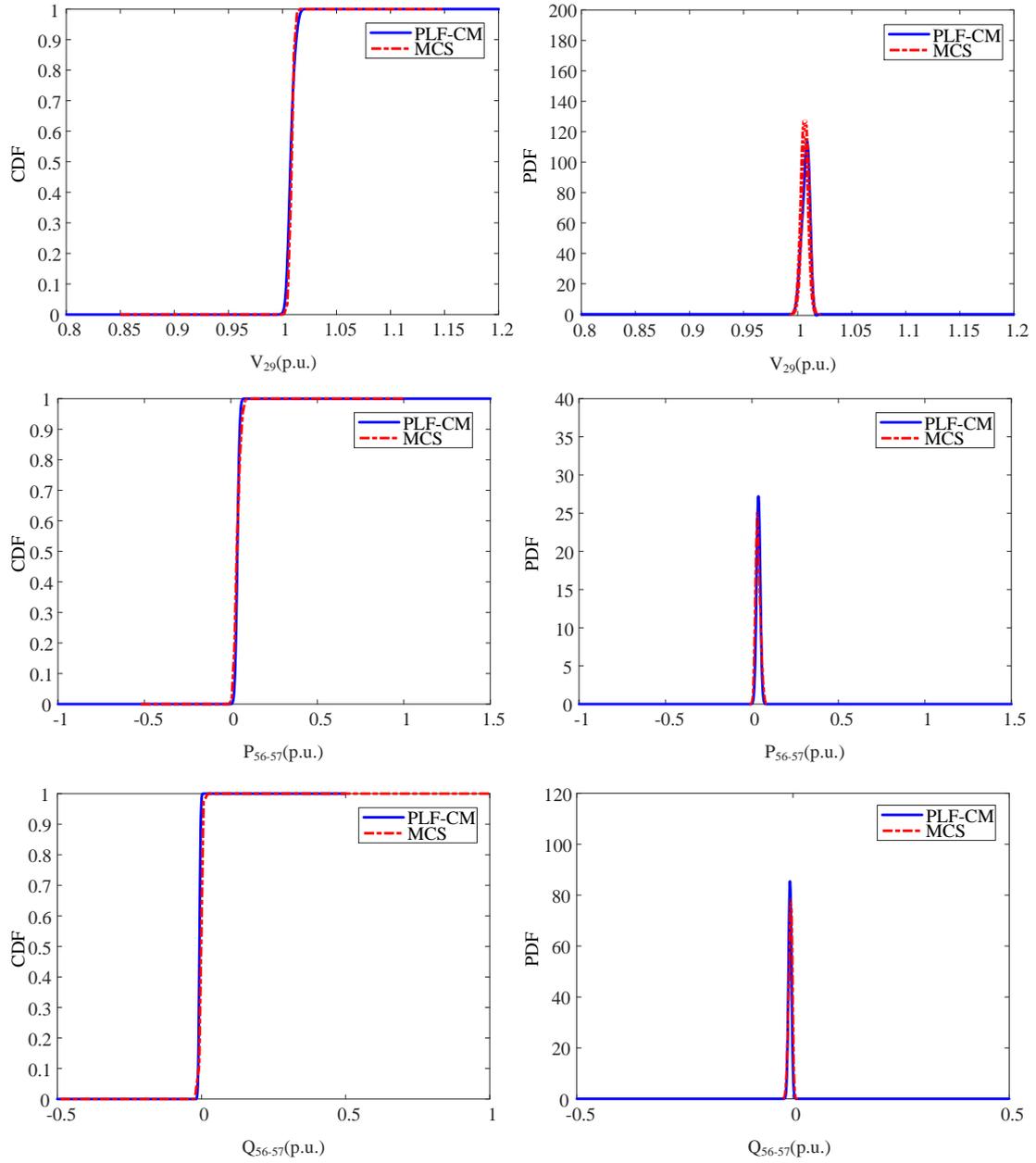

**FB. 3 PDF and CDF of output variables**

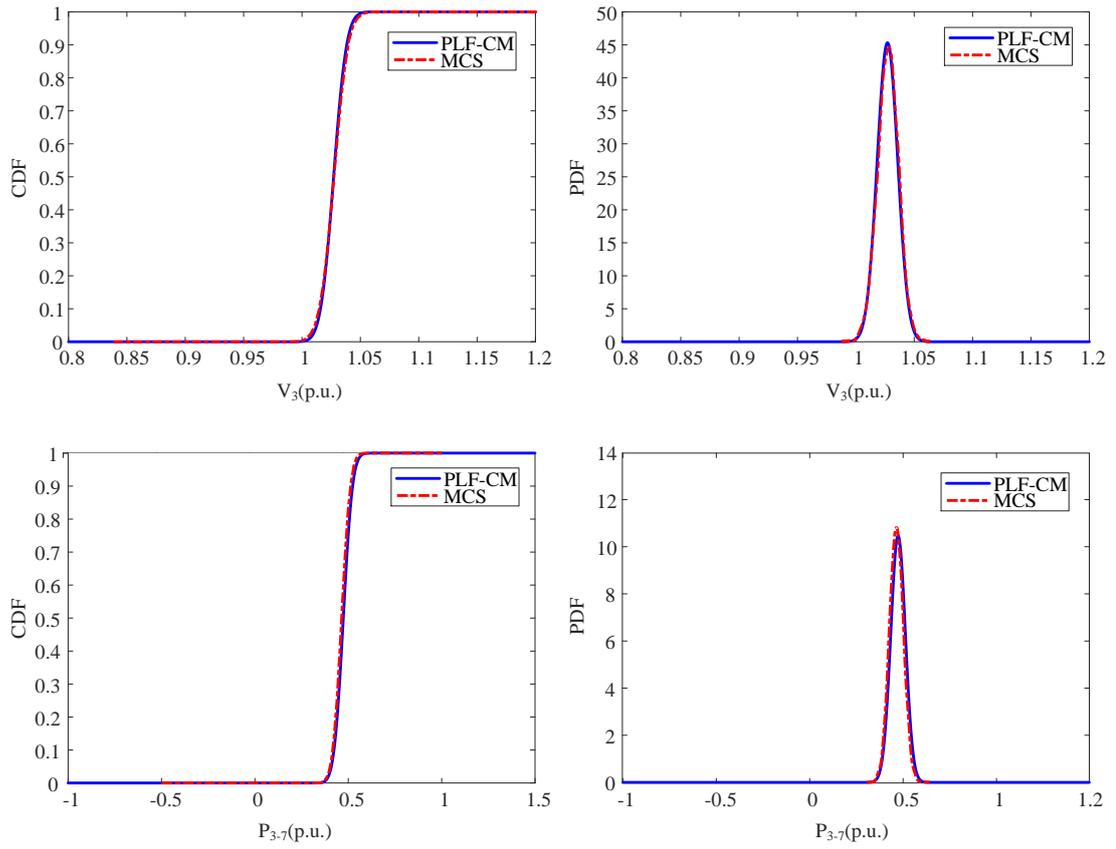

**FB. 4 PDF and CDF of output variables**

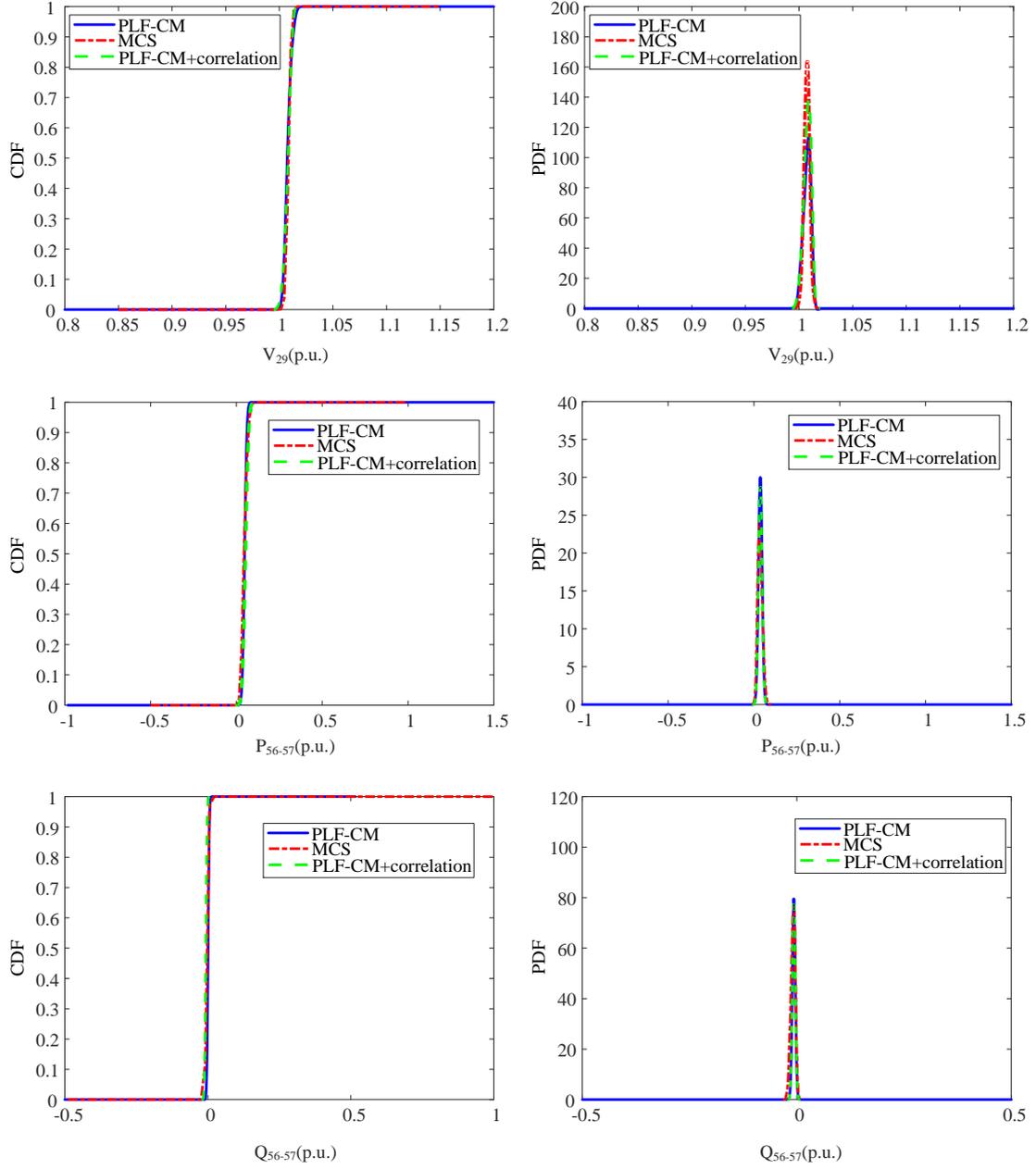

**FB. 5 PDF and CDF of output variables**

## Appendix C

To verify the universality of the method, a five-terminal DC grid was connected to the IEEE34 distribution network system, as shown in FC. 1. The parameters of the converter and the DC line are shown in TC. 1 and TC. 2. The control mode and scenario settings of the converter are shown in TC .3. The relative error and ARMS are also used to evaluate the effectiveness of the method. Due to the length of the article, this article only gives the results of the relative error index and the ARMS error index of the output variables, as shown in TC. 4 and TC. 5.

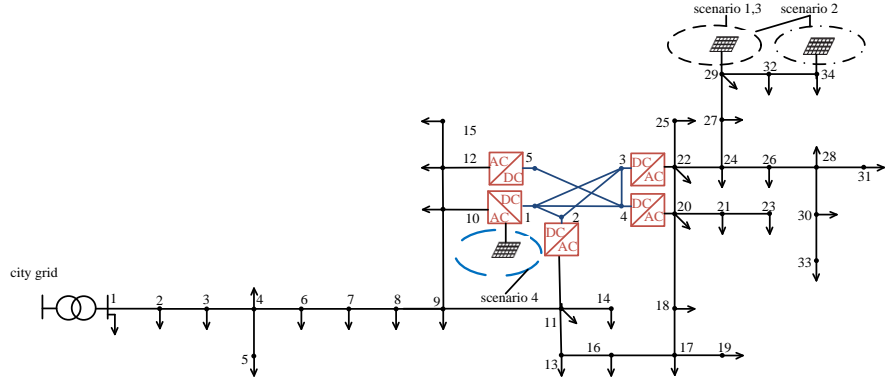

**FC. 1 Modified IEEE 34-node test feeder system**

**TC. 1 VSC parameters (p.u.)**

| VSC | Transformer parameters $Z_{tf}$ | Filter parameters $B_f$ | VSC parameters $Z_c$ |
|---|---|---|---|
| 1, 2, 3,4,5 | 0.00272+j0.203 | 0.04899 | 0.000181+j0.29743 |

**TC. 2 DC power grid line parameters (p.u.)**

| Head | End | Line resistance |
|---|---|---|
| 1 | 2 | 0.0304 |
| 1 | 3 | 0.0779 |
| 1 | 4 | 0.0779 |
| 2 | 3 | 0.1217 |
| 3 | 4 | 0.3115 |
| 4 | 5 | 0.0498 |

**TC. 3 VSC control parameters（p.u.）**

| Scenarios | Converter | Control mode | $u_{dc}$ | $U_s$ | $P_s$ | $Q_s$ |
|---|---|---|---|---|---|---|
| scenario 1,2 | 1 | $u$-$Q$ | 1.0 | \ | \ | 0.03 |
| | 2 | $P$-$Q$ | \ | \ | 0.05 | 0.03 |
| | 3 | $P$-$Q$ | \ | \ | -0.20 | 0.03 |
| | 4 | $P$-$Q$ | \ | \ | -0.16 | 0.03 |
| | 5 | $P$-$Q$ | \ | \ | 0.07 | 0.03 |
| scenario 3 | 1 | $u$-$Q$ | 1.0 | \ | \ | 0.03 |
| | 2 | $P$-$U$ | \ | 1.0 | 0.05 | \ |
| | 3 | $P$-$Q$ | \ | \ | -0.20 | 0.03 |
| | 4 | $P$-$Q$ | \ | \ | -0.16 | 0.03 |
| | 5 | $P$-$Q$ | \ | \ | 0.07 | 0.03 |
| scenario 4 | 1 | $f$-$U$ | 1.0 | \ | 1.0 | \ |
| | 2 | $P$-$Q$ | \ | \ | 0.05 | 0.03 |
| | 3 | $P$-$Q$ | \ | \ | -0.20 | 0.03 |
| | 4 | $P$-$Q$ | \ | \ | -0.16 | 0.03 |
| | 5 | $P$-$Q$ | \ | \ | 0.07 | 0.03 |

**TC. 4 Relative error indicators of output variables**

| Scenarios | Output variable | $\varepsilon^{\gamma}_{\mu}/10^{-2}$ | | $\varepsilon^{\gamma}_{\sigma}/10^{-2}$ | |
|---|---|---|---|---|---|
| | | $\varepsilon^{\gamma}_{\mu.\text{mean}}$ | $\varepsilon^{\gamma}_{\mu.\text{max}}$ | $\varepsilon^{\gamma}_{\sigma.\text{mean}}$ | $\varepsilon^{\gamma}_{\sigma.\text{max}}$ |
| | $U$ | 1.982 | 3.099 | 1.381 | 3.238 |
| scenario 1 | P | 1.133 | 3.499 | 1.768 | 4.589 |
| | Q | 1.067 | 2.677 | 3.677 | 4.848 |
| scenario 2 | U | 1.614 | 2.602 | 4.501 | 5.347 |
| | P | 1.622 | 4.544 | 1.839 | 4.612 |
| | Q | 1.262 | 4.050 | 1.565 | 4.245 |
| scenario 3 | U | 0.844 | 1.603 | 2.087 | 2.542 |
| | P | 1.086 | 3.379 | 1.107 | 2.854 |
| | Q | 1.257 | 4.475 | 1.222 | 2.594 |
| scenario 4 | $u_{dc}$ | 0.922 | 2.453 | 1.103 | 3.022 |
| | $P_{dc}$ | 0.750 | 1.220 | 1.402 | 2.501 |

**TC. 5 ARMS indicators of output variables**

| Scenarios | Output variable | $\varepsilon^\gamma_{mean}/10^{-2}$ | $\varepsilon^\gamma_{max}/10^{-2}$ |
|---|---|---|---|
| scenario 1 | U | 0.74 | 1.62 |
| | P | 0.88 | 2.21 |
| | Q | 0.18 | 0.32 |
| scenario 2 | U | 0.73 | 1.60 |
| | P | 0.87 | 1.87 |
| | Q | 0.18 | 0.33 |
| scenario 3 | U | 0.90 | 2.17 |
| | P | 0.89 | 2.22 |
| | Q | 0.25 | 0.64 |
| scenario 4 | $u_{dc}$ | 0.88 | 1.75 |
| | $P_{dc}$ | 0.21 | 0.75 |

It can be seen from TC. 4 and TC. 5 that for different Scenarios, the maximum values of $\varepsilon^\gamma_{\mu.mean}$, $\varepsilon^\gamma_{\mu.max}$, $\varepsilon^\gamma_{\sigma.mean}$ and $\varepsilon^\gamma_{\sigma.max}$ are 1.982%, 4.544%, 4.501%, and 5.348%, respectively, and the maximum values of $\varepsilon^\gamma_{mean}$ and $\varepsilon^\gamma_{max}$ are 0.9% and 2.22%, respectively.

## Appendix D

The resistance R is used to simulate the sum of active loss of VSC and converter reactor, this method simplifies the converter loss greatly, which is also adopted in this paper.

In general, the power instruction of the converter station controller refers to the ac power. However, in the calculation of the power flow of the DC grid, the known node power is the DC power injected by the converter station into the DC grid, so the loss of the converter station needs to be estimated. Assuming that the converter station works in the rectification state (it can also be deduced by the same reasoning in the inverter side), the relationship between ac power, DC power and loss is as follows:

$$P_{ac} = P_{dc} + P_{vsc-loss} \quad (1)$$

The unitary value form of Equation (1) is:

$$P_{ac,pu} = P_{dc,pu} + r_{pu}\left(i_{d,pu}^2 + i_{q,pu}^2\right) \quad (2)$$

Where $i_{d,pu}$, $i_{q,pu}$ ——d and q axis unit values components of the grid current;

$r_{pu}$ ——The unit value of the equivalent loss resistance R of the converter

Assuming that the AC system side voltage is stable near the rated value, let axis D coincide with the grid voltage vector, so that axis D coincides with the grid voltage

vector, then

$$P_{ac,pu} = 1.5 u_{sd,pu} i_{d,pu} = 1.5 i_{d,pu} \quad (3)$$

When the VSC adopts unit power factor control, $i_{q,pu}=0$; Substitute equation (3) into Equation (2), then：

$$r_{pu} = \frac{(P_{ac,pu} - P_{dc,pu})}{\left(\dfrac{2}{3} P_{ac,pu}\right)^2} \quad (4)$$

$P_{dc,pu}$ can be obtained in the calculation of dc power flow, $P_{ac,pu}$ is the fixed value of control. Therefore, the equivalent resistance of the converter loss $r_{pu}$ can be estimated，and combine it with the resistors of the converter reactor and transformer.

This method avoids repeated processing of network loss in the process of each iteration, and has very little error compared with the quadratic function method (which requires two variables of power and voltage), which is not elaborated in the paper due to limited space.